\documentclass[aps,prb,twocolumn,superscriptaddress]{revtex4}
\usepackage{graphicx}
\usepackage{graphicx,color}
\usepackage{epstopdf}
\usepackage{amsfonts}
\usepackage{amsmath,amssymb,mathrsfs}
\usepackage{bm}
\usepackage{float}
\usepackage{pst-grad}
\usepackage{dcolumn}
\usepackage{units}
\usepackage{nicefrac}
\usepackage{array}
\usepackage{booktabs}
\usepackage{hhline}
\usepackage{subfigure}

\newcommand\Tstrut{\rule{0pt}{2.5ex}}

\begin{document}

\title{Measurement of the $B_{1g}$ and $B_{2g}$ components of the elastoresistivity tensor for tetragonal materials via transverse resistivity configurations}

\author{M. C. Shapiro}
\email[Corresponding author: ]{maxshaps@stanford.edu}
\affiliation{Stanford Institute for Materials and Energy Sciences, SLAC National Accelerator Laboratory,\\ 2575 Sand Hill Road, Menlo Park, California 94025, USA} 
\affiliation{Geballe Laboratory for Advanced Materials and Department of Applied Physics, Stanford University, Stanford, California 94305, USA} 
\author{A. T. Hristov}
\affiliation{Stanford Institute for Materials and Energy Sciences, SLAC National Accelerator Laboratory,\\ 2575 Sand Hill Road, Menlo Park, California 94025, USA} 
\affiliation{Geballe Laboratory for Advanced Materials and Department of Physics, Stanford University, Stanford, California 94305, USA}
\author{J. C. Palmstrom}
\affiliation{Stanford Institute for Materials and Energy Sciences, SLAC National Accelerator Laboratory,\\ 2575 Sand Hill Road, Menlo Park, California 94025, USA} 
\affiliation{Geballe Laboratory for Advanced Materials and Department of Applied Physics, Stanford University, Stanford, California 94305, USA}
\author{Jiun-Haw Chu}
\affiliation{Stanford Institute for Materials and Energy Sciences, SLAC National Accelerator Laboratory,\\ 2575 Sand Hill Road, Menlo Park, California 94025, USA} 
\affiliation{Geballe Laboratory for Advanced Materials and Department of Applied Physics, Stanford University, Stanford, California 94305, USA}
\author{I. R. Fisher}
\affiliation{Stanford Institute for Materials and Energy Sciences, SLAC National Accelerator Laboratory,\\ 2575 Sand Hill Road, Menlo Park, California 94025, USA} 
\affiliation{Geballe Laboratory for Advanced Materials and Department of Applied Physics, Stanford University, Stanford, California 94305, USA}

\begin{abstract}

The elastoresistivity tensor $m_{ij,kl}$ relates changes in resistivity to strains experienced by a material. As a fourth-rank tensor, it contains considerably more information about the material than the simpler (second-rank) resistivity tensor; in particular, for a tetragonal material, the $B_{1g}$ and $B_{2g}$ components of the elastoresistivity tensor ($m_{xx,xx}-m_{xx,yy}$ and $2m_{xy,xy}$, respectively) can be related to its nematic susceptibility.\cite{kuo_2013,shapiro_2015}  Previous experimental probes of this quantity have focused exclusively on differential longitudinal elastoresistance measurements,\cite{kuo_2013,chu_2012,kuo_2014,riggs_2015} which determine the induced resistivity anisotropy arising from anisotropic in-plane strain based on the difference of two longitudinal resistivity measurements. Here we describe a complementary technique based on \textit{transverse} elastoresistance measurements. This new approach is advantageous because it directly determines the strain-induced resistivity anisotropy from a single transverse measurement. To demonstrate the efficacy of this new experimental protocol, we present transverse elastoresistance measurements of the $2m_{xy,xy}$ elastoresistivity coefficient of BaFe$_2$As$_2$, a representative iron-pnictide that has previously been characterized via differential longitudinal elastoresistance measurements.

\end{abstract}
\maketitle

\section{Introduction}

\begin{figure}[t!]
\includegraphics[clip=true, width=\columnwidth]{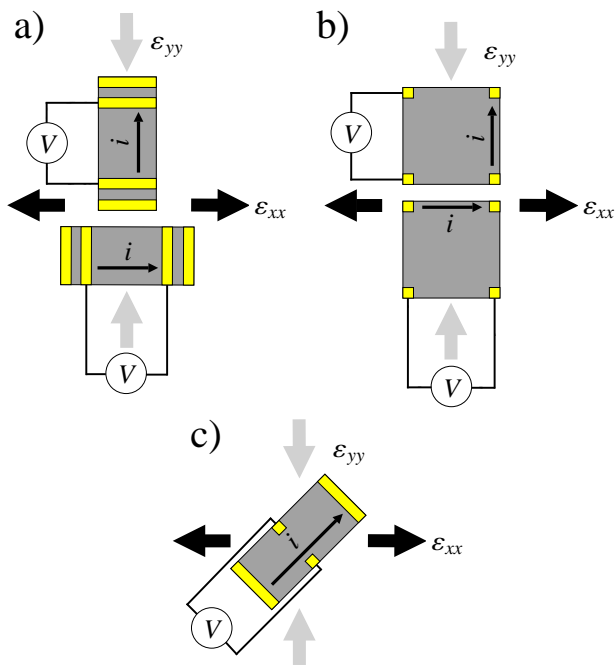}
\caption{Schematic diagrams illustrating three different methods that can be used to measure the $B_{1g}$ or $B_{2g}$ elastoresistivity coefficients of a tetragonal material. The appropriate orientation of the principal crystal axes with respect to the current $\boldsymbol{i}$ and the strains $\epsilon_{xx}$ (horizontal black arrows) and $\epsilon_{yy}$ (vertical gray arrows) for the two irreducible components is discussed in Sections \ref{sec:difflong} and \ref{sec:transverse} of the main text. Gray bars represent single crystal samples and yellow regions depict electrical contacts used for the transport measurements; current paths and schematic voltmeters are also indicated. The samples are caused to experience an induced anisotropic strain by some external means and the strains $\epsilon_{xx}$ and $\epsilon_{yy}$ are separately determined.  In the configuration shown in panel (a), a standard four-point contact geometry is used to measure the differential longitudinal elastoresistivity (i.e., $\left(\nicefrac{\Delta \rho}{\rho}\right)_{xx} - \left(\nicefrac{\Delta \rho}{\rho}\right)_{yy}$). In panel (b), a modified Montgomery geometry is used to measure the same quantities with a single sample; however, two measurement configurations are still required to extract $\left(\nicefrac{\Delta \rho}{\rho}\right)_{xx}$ and $\left(\nicefrac{\Delta \rho}{\rho}\right)_{yy}$, as illustrated by the two schematic diagrams. In both cases, the relevant elastoresistivity coefficients are determined from the difference of the two longitudinal measurements.  As described in the main text, these configurations have some practical drawbacks; in particular, one infers a (potentially small) resistive anisotropy from the difference of two (potentially larger) longitudinal resistivity measurements.  In this paper we describe an alternative \textit{transverse} (i.e., $\left(\nicefrac{\Delta \rho}{\rho}\right)_{xy}$) measurement that can be performed on one single crystal (depicted in panel (c)) that yields the same elastoresistivity coefficients by directly determining the resistivity anisotropy from a single measurement.}
\label{fig:difflongvstrans}
\end{figure}

Resistivity measurements are employed extensively in the field of strongly correlated electron systems (SCES).  Since transport properties are determined by the electronic dynamics at the Fermi level, resistivity is often extremely sensitive to Fermi surface changes and electronically-driven phase transitions; however, since resistivity is a second-rank tensor, transport measurements are generically limited in their ability to identify the \textit{symmetry} properties of the underlying order.  In contrast, the elastoresistivity (a fourth-rank tensor defined as the strain derivative of the resistivity) can convey additional information about directional anisotropies and broken point group symmetries which might more subtly manifest in the resistivity itself.\cite{shapiro_2015,hlobil_2015}  Furthermore, since electron-lattice coupling in SCES is often large, the order parameter characterizing an electronically-driven phase transition in these materials is often strongly tuned by strain and strongly reflected in transport; the coefficients in the elastoresistivity tensor are then likely to be large, making elastoresistivity very promising from an experimental perspective.  Although elastoresistance measurements have been applied to semiconductors,\cite{sun_2010} this physical quantity has only recently been measured in the context of SCES;\cite{kuo_2013,shapiro_2015,chu_2012,kuo_2014,riggs_2015,watson_2015,kuo_2015,tanatar_2015} in both cases, however, measurements have been confined to \textit{longitudinal} geometries (Figure~\ref{fig:difflongvstrans} (a) and (b)), and the wider class of \textit{transverse} (Figure~\ref{fig:difflongvstrans} (c)) measurements (which are the subject of this paper) have not been investigated.

For a tetragonal material, the $B_{1g}$ and $B_{2g}$ components of the elastoresistivity tensor characterize the material's linear response to the anisotropic strains $\epsilon_{xx} - \epsilon_{yy}$ and $\epsilon_{xy}$, respectively. These two components of the elastoresistivity tensor directly connect to the nematic susceptibility for the same two symmetry channels, $\chi_{_{B_{1g}}}$ and $\chi_{_{B_{2g}}}$.\cite{kuo_2013, shapiro_2015} We have recently shown how these coefficients can be determined from differential longitudinal elastoresistance measurements and have used this technique to investigate a series of materials which exhibit electronic nematic instabilities.\cite{kuo_2013,chu_2012,kuo_2014,riggs_2015,kuo_2015}

Anisotropic strain can be achieved by gluing crystals to the side surface of a piezoelectric lead zirconate titanate (PZT) stack with a strain-transmitting epoxy.  In this implementation, the crystals are mechanically coupled to and hence deform with the PZT, which expands (contracts) along its poling direction (perpendicular to its poling direction) upon application of a positive external voltage. Longitudinal resistances are then measured while the strain is varied and the differential longitudinal elastoresistance determined from the difference of the two measurements.  In the original realization of this experiment, two separate transport bars were used in order to separately determine the longitudinal elastoresistivities $\left(\nicefrac{\Delta \rho}{\rho}\right)_{xx}$ and $\left(\nicefrac{\Delta \rho}{\rho}\right)_{yy}$ (illustrated schematically in Figure~\ref{fig:difflongvstrans} (a)).  While these measurements unambiguously identified a divergence of the nematic susceptibility in the $B_{2g}$ symmetry channel for the iron-based superconductors,\cite{kuo_2013} nevertheless this specific experimental configuration leads to several experimental concerns.  In particular, the technique relies upon equal strain transmission for the two samples used in the differential measurement (which might be difficult to realize in practice, in part due to geometric factors and in part due to differences in the adhesion of the two samples to the PZT stack).  Expressed in the context of group theory, such nonidealities (which we describe in greater detail in Section \ref{sec:difflong}) admix elastoresistivity coefficients with an $A_{1g}$ character (i.e., isotropic in-plane, or symmetric with respect to rotation about the $c$-axis), potentially affecting the determination of the associated elastoresistivity coefficients in other symmetry channels.

To avoid the possible contamination of isotropic strain that can manifest in a differential longitudinal elastoresistivity measurement, it is preferable to extract the differential elastoresistivity from a measurement performed on just one single crystal sample that is held under conditions of anisotropic strain.  One such method is to use the modified Montgomery technique to measure the induced anisotropy in the longitudinal resistivity of a square shape sample (Figure~\ref{fig:difflongvstrans} (b)).  We recently applied such a technique to measure the differential elastoresistance of several families of iron-based superconductors.\cite{kuo_2016}  While this technique obviates concerns over strain transmission to the sample, nevertheless it still requires separate measurement of two (potentially large) longitudinal resistivities as a function of strain, the (potentially small) difference of which yields the desired $B_{1g}$ or $B_{2g}$ components of the elastoresistivity tensor. Ideally, one would determine this difference directly.

We note that it is indeed possible to measure the induced resistivity anisotropy (and hence the $B_{1g}$ and $B_{2g}$ components of the elastoresistivity tensor) from a single measurement.  In particular, we note that a tetragonal material which undergoes an orthorhombic distortion by breaking symmetry about its $\sigma_x$ and $\sigma_y$ mirror planes (i.e., undergoes a $B_{2g}$ distortion in which the in-plane square lattice deforms into a parallelogram) acquires finite off-diagonal terms in the resistivity tensor ($\rho_{xy}$ and $\rho_{yx}$) which are proportional to the amount of orthorhombicity.  Hence, one can obtain the same information \textit{from a single measurement} of the \textit{transverse} elastoresistivity (Figure~\ref{fig:difflongvstrans} (c)).  The primary advantage of the transverse method is that it directly measures the associated resistive anisotropy from a single measurement of a single sample. Furthermore, by symmetry, the measured quantity cannot be affected by isotropic strain in the linear regime.

In this manuscript, we propose and demonstrate a new method for probing the nematic susceptibility in the $B_{2g}$ channel $\chi_{_{B_{2g}}}$ based on measuring the transverse elastoresistivity $\left(\nicefrac{\Delta \rho}{\rho}\right)_{xy}$. One common problem that can arise with measurements of a transverse resistivity is $\rho_{xx}$ contamination in a nominal $\rho_{xy}$ measurement due to contact misalignment, and so we also provide a practical means for subtracting such contamination.  Since transverse elastoresistivity measurements have to date neither been discussed nor measured, we provide here a detailed description of the relevant tensor quantities and a suitable technique that enables such a measurement.

We proceed by first describing appropriate coordinate frames and associated transformations of the elastoresistivity tensor, necessary for the subsequent discussion.  We then explain the various configurations for measuring the corresponding elastoresistivity coefficients, along the way characterizing certain forms of experimental error.  We conclude by presenting $2m_{xy,xy}$ data acquired via the new method for the representative iron-pnictide BaFe$_2$As$_2$, which was chosen since it has previously been well-characterized by differential longitudinal measurements\cite{chu_2012,kuo_2013,kuo_2014} and has a large elastoresistive response.  The temperature dependence of the elastoresistivity coefficients as observed by the transverse method agree with the earlier differential measurements, revealing a nematic instability in the $B_{2g}$ symmetry channel.  Similar to our earlier differential longitudinal measurements, anisotropic strain for the transverse elastoresistance measurements was achieved by gluing the sample to the surface of a piezoelectric PZT stack; however, we note that the proposed technique does not rely on this specific realization, and alternative methods to strain the sample can be readily envisioned.

\section{Coordinate Frames and the Elastoresistivity Tensor}

As a consequence of strains experienced by a material, terms in the resistivity tensor $\rho_{ij}$ acquire a strain-induced change

\begin{equation}
\label{eq:1}
\Delta \rho_{ij}(\boldsymbol H) \equiv \rho_{ij}(\boldsymbol H,\hat{\epsilon}) - \rho_{ij}(\boldsymbol H,\hat{\epsilon}=\hat{0}).
\end{equation}

\noindent The elastoresistivity $m_{ij,kl}(\boldsymbol H )$ is a fourth-rank tensor that linearly relates the (normalized) strain-induced resistivity change $\left(\nicefrac{\Delta \rho}{\rho}\right)_{ij}(\boldsymbol H)$ and the strain $\epsilon_{kl}$ according to

\begin{equation}
\label{eq:2}
\left(\nicefrac{\Delta \rho}{\rho}\right)_{ij}(\boldsymbol H) \equiv m_{ij,kl}(\boldsymbol H ) \epsilon_{kl},
\end{equation}

\noindent where we choose to represent the second-rank tensors $\left(\nicefrac{\Delta \rho}{\rho}\right)_{ij}(\boldsymbol H)$ and $\epsilon_{kl}$ as the column vectors

\begin{equation}
\label{eq:3}
\hspace{-2mm} \left(\nicefrac{\Delta \rho}{\rho}\right)_{ij}(\boldsymbol H) = \begin{pmatrix} \left(\nicefrac{\Delta \rho}{\rho}\right)_{xx}(\boldsymbol H) \\ \left(\nicefrac{\Delta \rho}{\rho}\right)_{yy}(\boldsymbol H) \\ \left(\nicefrac{\Delta \rho}{\rho}\right)_{zz}(\boldsymbol H) \\ \left(\nicefrac{\Delta \rho}{\rho}\right)_{yz}(\boldsymbol H) \\ \left(\nicefrac{\Delta \rho}{\rho}\right)_{zy}(\boldsymbol H) \\ \left(\nicefrac{\Delta \rho}{\rho}\right)_{zx}(\boldsymbol H) \\ \left(\nicefrac{\Delta \rho}{\rho}\right)_{xz}(\boldsymbol H) \\ \left(\nicefrac{\Delta \rho}{\rho}\right)_{xy}(\boldsymbol H) \\ \left(\nicefrac{\Delta \rho}{\rho}\right)_{yx}(\boldsymbol H) \end{pmatrix} \hspace{1mm} \textrm{and} \hspace{2mm} \epsilon_{kl} = \begin{pmatrix} \epsilon_{xx} \\ \epsilon_{yy} \\ \epsilon_{zz} \\ \epsilon_{yz} \\ \epsilon_{zy} \\ \epsilon_{zx} \\ \epsilon_{xz} \\ \epsilon_{xy} \\ \epsilon_{yx} \end{pmatrix}
\end{equation}
\vspace{\baselineskip}

\noindent in order to represent $m_{ij,kl}(\boldsymbol H)$ as a $9 \times 9$ matrix. The appropriate normalization scheme is given by\cite{shapiro_2015}

\begin{equation}
\label{eq:4}
(\nicefrac{\Delta \rho}{\rho})_{ij}(\boldsymbol H) \equiv (\nicefrac{\Delta \rho_{ij}(\boldsymbol H)}{\sqrt{\rho_{ii}(\boldsymbol H)}\sqrt{\rho_{jj}(\boldsymbol H)}}).
\end{equation}

\noindent Because of Onsager's relation,\cite{onsager_1931} the resistivity tensor is not in general symmetric in the presence of a magnetic field and so we avoid usage of the compactified Voigt notation in order to present a generalized description appropriate for finite $\boldsymbol H$.  The point group symmetry of the crystal lattice constrains the number of independent nonzero coefficients in the elastoresistivity tensor; for example, the elastoresistivity tensor for the specific case of the $D_{4h}$ point group (appropriate for BaFe$_2$As$_2$ and derived elsewhere\cite{shapiro_2015}) is given in Appendix \ref{appendix:tensortransform}.

In labeling the elastoresistivity coefficients by spatial coordinates, we have implicitly assumed a Cartesian system referenced to the crystal itself and defined by its primitive lattice vectors.  We refer to this reference frame as the ``crystal frame'' and denote it by unprimed $x$, $y$, and $z$ axes.  In order to extract symmetry information about the crystal, one is generally concerned with measured quantities in the crystal frame.  We consider an experiment in which the crystal experiences a purely normal (i.e., shear-free) homogeneous strain in a given Cartesian frame of reference defined by $x'$, $y'$, and $z'$ basis vectors.  For example, this could be realized with a piezoelectric PZT stack, where the basis vectors are defined by the lateral dimensions of the stack.  We refer to this reference frame as the ``normal strain frame'', which (by choice) shares a mutual $z / z'$ axis with the crystal frame but is oriented at an in-plane angle $\phi$ relative to the primitive axes of the crystal frame (i.e., $\hat{x} \cdot \hat{x}' = \hat{y} \cdot \hat{y}' = \cos\phi$, where $\phi$ is positive when the crystal frame is oriented counterclockwise relative to the normal strain frame).  The relative angle $\phi$ reflects our freedom to strain the crystal along arbitrary directions relative to the primitive crystal cell. 

\begin{figure}
\includegraphics[clip=true, width=\columnwidth]{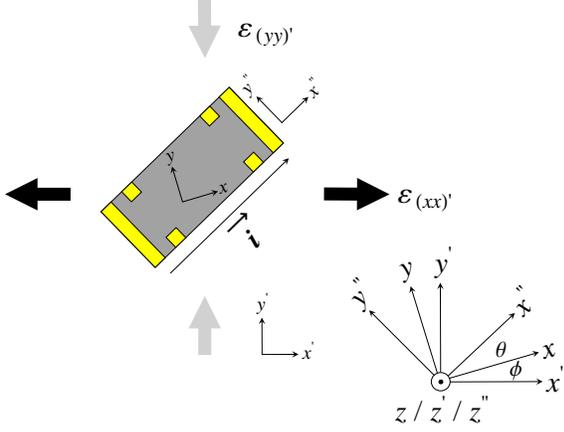}
\caption{Schematic diagram illustrating the relative orientations of the crystal (unprimed), normal strain (primed), and transport (double primed) coordinate frames.  The crystal and normal strain frames are related by a relative angle $\phi$ about their mutual $z / z'$ axis, while the crystal and transport frames are related by a relative angle $\theta$ about their mutual $z / z''$ axis.}
\label{fig:coordinates}
\end{figure}

Additionally, when we perform an in-plane resistivity measurement, we have the freedom to direct the current along an arbitrary in-plane direction with respect to the crystal axes.  We define this ``transport frame'' by double-primed Cartesian vectors $x''$, $y''$, and $z''$; $x''$ is the direction in which the current is sourced, $y''$ is the in-plane direction perpendicular to $x''$, and $z''$ is the out-of-plane direction perpendicular to $x''$.  The transport frame shares a mutual $z / z''$ axis with the crystal frame but is oriented at an in-plane angle $\theta$ relative to it (i.e., $\hat{x} \cdot \hat{x}'' = \hat{y} \cdot \hat{y}'' = \cos\theta$, where $\theta$ is positive when the crystal frame is oriented counterclockwise relative to the current frame).  The relative orientation of the three coordinate frames is depicted in Figure \ref{fig:coordinates}.

When performing an in-plane elastoresistivity measurement, the normalized changes in resistivity $\left(\nicefrac{\Delta \rho}{\rho}\right)_{(xx)''}(H_z)$, $\left(\nicefrac{\Delta \rho}{\rho}\right)_{(yy)''}(H_z)$, $\left(\nicefrac{\Delta \rho}{\rho}\right)_{(xy)''}(H_z)$, and $\left(\nicefrac{\Delta \rho}{\rho}\right)_{(yx)''}(H_z)$ are measured in the transport frame (which is rotated relative to the crystal frame by an angle $\theta$), while the strains $\epsilon_{(xx)'}$, $\epsilon_{(yy)'}$ and $\epsilon_{(zz)'}$ are measured in the normal strain frame (which is rotated relative to the crystal frame by an angle $-\phi$); they are related by means of appropriately transformed elastoresistivity coefficients according to

\begin{align}
\label{eq:5}
(\nicefrac{\Delta \rho}{\rho})_{(ij)''} &= \hat{\alpha}_{\theta} (\nicefrac{\Delta \rho}{\rho})_{ij} = \hat{\alpha}_{\theta} m_{ij,kl} \epsilon_{kl} \\
&= \hat{\alpha}_{\theta} m_{ij,kl} \hat{\alpha}_{\phi} \epsilon_{(kl)'} \nonumber \\
&\equiv m_{(ij)'',(kl)'} \epsilon_{(kl)'} \nonumber,
\end{align}

\noindent where the $\hat{\alpha}_{\phi}$, $\hat{\alpha}_{\theta}$ are suitable transformation matrices given in Appendix \ref{appendix:tensortransform} and the subscripts in the elastoresistivity coefficients denote that the strains are measured in the normal strain frame (primes) while the normalized changes in resistivity are measured in the transport frame (double primes).

\section{Differential Longitudinal Configuration for Probing Nematic Susceptibility in $D_{4h}$}\label{sec:difflong}

\subsection{Ideal Configuration}

\begin{figure}
\centering
\includegraphics[width=\columnwidth]{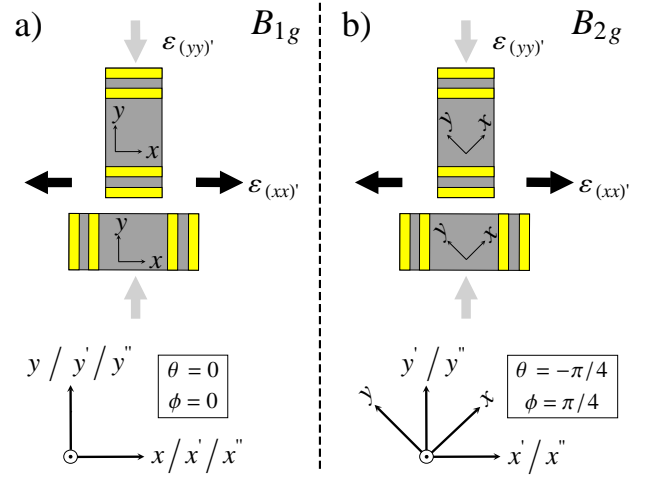}
\caption{Longitudinal elastoresistivity configurations for extracting (a) $m_{xx,xx} - m_{xx,yy}$ (with $(\theta, \phi) = (0,0)$) and (b) $2m_{xy,xy}$ (with $(\theta, \phi) = (-\pi/4, \pi/4)$), which characterize the $B_{1g}$ and $B_{2g}$ irreducible representations of $m_{ij,kl}$ in $D_{4h}$.  In (a), one measures the differential longitudinal resistive response to strain $\left(\nicefrac{\Delta \rho}{\rho}\right)_{(xx)''} - \left(\nicefrac{\Delta \rho}{\rho}\right)_{(yy)''}$ to a strain $\epsilon_{(xx)'}-\epsilon_{(yy)'}$ with the transport, crystal, and normal strain frames all coincident; the differential longitudinal elastoresistivity then yields the elastoresistivity coefficients $m_{xx,xx} - m_{xx,yy}$.  In (b), one measures the differential resistive response to strain $\left(\nicefrac{\Delta \rho}{\rho}\right)_{(xx)''} - \left(\nicefrac{\Delta \rho}{\rho}\right)_{(yy)''}$ to a strain $\epsilon_{(xx)'}-\epsilon_{(yy)'}$ with the crystal frame oriented at the angles $(\theta, \phi) = (-\pi/4, \pi/4)$ relative to the transport and normal strain frames; the differential longitudinal elastoresistivity then yields the elastoresistivity coefficient $2m_{xy,xy}$.}
\label{fig:symmetryconfiga}
\end{figure}

The elastoresistivity tensor takes a particularly simple form when decomposed in terms of its irreducible representations (as determined by the point group symmetry of the crystal lattice).  Such a decomposition motivates making specific combinations of elastoresistance measurements in order to isolate particular elastoresistivity coefficients in the same symmetry class.  For example, for the $D_{4h}$ point group, the normalized resistivity changes in the $B_{1g}$ and $B_{2g}$ irreducible representations are proportional to the corresponding elastoresistivity coefficients ($m_{xx,xx} - m_{xx,yy}$ and $2m_{xy,xy}$, respectively) in the same symmetry channel and are related to the nematic susceptibilities $\chi_{_{B_{1g}}}$ and $\chi_{_{B_{2g}}}$ in those irreps.\cite{kuo_2013,riggs_2015,shapiro_2015}  There are, however, different experimental geometries that can be used to extract these combinations of coefficients.  Previously, we have shown how a differential longitudinal elastoresistance measurement can be used to measure $m_{xx,xx} - m_{xx,yy}$ and $2m_{xy,xy}$, which is illustrated schematically in Figure~\ref{fig:symmetryconfiga}.  By taking the symmetry-motivated combination $\left(\nicefrac{\Delta \rho}{\rho}\right)_{(xx)''} - \left(\nicefrac{\Delta \rho}{\rho}\right)_{(yy)''}$ and expressing it in terms of the elastoresistivity coefficients in the crystal frame (Appendix \ref{appendix:tensortransform}), we find that

\begin{align}
\label{eq:6}
\left(\nicefrac{\Delta \rho}{\rho}\right)&_{(xx)''}(H_z) - \left(\nicefrac{\Delta \rho}{\rho}\right)_{(yy)''}(H_z) = \\
& \Big[ \epsilon_{(xx)'} - \epsilon_{(yy)'} \Big] \cdot \Big[ (m_{xx,xx} - m_{xx,yy}) \cos(2\theta) \cos(2\phi) \nonumber \\
& \hspace{2.75cm} - 2m_{xy,xy} \sin(2\theta) \sin(2\phi) \Big] \nonumber.
\end{align}

\noindent Unsurprisingly, despite the fact that each crystal experiences $\epsilon_{(zz)'}$ strain (and so the individual transport measurements experience the effects of strain in the $A_{1g}$ symmetry channel in addition to the $B_{1g}$ or $B_{2g}$ channels), the quantity $\left(\nicefrac{\Delta \rho}{\rho}\right)_{(xx)''} - \left(\nicefrac{\Delta \rho}{\rho}\right)_{(yy)''}$ is unaffected by such strains since they are of a different symmetry class. Equivalently, the effects of rotationally invariant strains are subtracted out in taking the $B_{1g}$ or $B_{2g}$ combination, as we originally noted.\cite{kuo_2013}  However, arbitrary in-plane rotations are not symmetry elements of $D_{4h}$ (only the discrete $\pi/2$ rotations about the mutual $z / z' / z''$ axis are symmetries of the point group), and so rotating the crystal frame by an arbitrary angle $\theta$ relative to the transport frame and/or by an arbitrary angle $\phi$ relative to the strain frame can mix $B_{1g}$ ($m_{xx,xx} - m_{xx,yy}$) and $B_{2g}$ ($2m_{xy,xy}$) quantities (measured in the crystal frame) into each other.  For certain high symmetry directions of the current and normal strains relative to the crystal axes, though, one can isolate the $B_{1g}$ and $B_{2g}$ coefficients and hence infer the behavior of the corresponding susceptibilities in those symmetry channels.  The high symmetry configuration for $m_{xx,xx} - m_{xx,yy}$ ($\propto \chi_{_{B_{1g}}}$) is $(\theta, \phi) = (0,0)$ (i.e., the transport, current, and normal strain frames are all coincident), while the high symmetry configuration for $2m_{xy,xy}$ ($\propto \chi_{_{B_{2g}}}$) is $(\theta, \phi) = (-\pi/4,\pi/4)$ (i.e., the crystal frame is oriented at $\pi/4$ radians relative to the transport and normal strain frames); these arrangements are depicted in Figure~\ref{fig:symmetryconfiga} and can be confirmed with \eqref{eq:6}.  This was precisely the configuration used in our initial measurements of the elastoresistance of iron-based\cite{chu_2012,kuo_2013} and heavy fermion\cite{riggs_2015} superconductors.  The same information can also be extracted from a modified Montgomery geometry.\cite{kuo_2016}

\subsection{Sources of Error}\label{sec:difflongerrors}

There are several sources of systematic errors in the differential longitudinal measurement configuration, all of which merit a brief comment.  The primary reason for doing so in the context of this paper is then to motivate the alternative transverse measurement configuration which does not suffer some of these drawbacks.  Here we focus specifically on errors associated with the standard four-contact geometry (Figure~\ref{fig:difflongvstrans} (a)), but a similar analysis could be applied to the modified Montgomery technique.

\hfill \break

\noindent \textbf{(a) \underline{Angular misalignment:}} In order to measure $m_{xx,xx} - m_{xx,yy}$ or $2m_{xy,xy}$, crystals should be oriented such that $(\theta, \phi) = (0,0)$ and $(\theta, \phi) = (-\pi/4, \pi/4)$, respectively.  In practice, misalignment will occur which will affect the measured elastoresistivity coefficients.  Within the present formalism, we can propagate this error to leading order; in this section, we quote the main results, referring the reader to Appendix~\ref{appendix:alignmenterrors} for the full derivation.

Suppose that in attempting to measure $\left(\nicefrac{\Delta \rho}{\rho}\right)_{(xx)''}$, we intended to orient the crystal in a high-symmetry configuration characterized by the angles $(\theta, \phi)$ but actually did so in a configuration given by $(\theta + \delta \theta_{xx}, \phi + \delta \phi_{xx})$.  To probe nematic susceptibility, we subtract $\left(\nicefrac{\Delta \rho}{\rho}\right)_{(yy)''}$ from  $\left(\nicefrac{\Delta \rho}{\rho}\right)_{(xx)''}$, which we also intend to be measured in a configuration $(\theta, \phi)$ but which may also be misaligned according to $(\theta + \delta \theta_{yy}, \phi + \delta \phi_{yy})$.  For full generality, we assume $\delta \theta_{xx} \neq \delta \theta_{yy}$ and $\delta \phi_{xx} \neq \delta \phi_{yy}$.  Expanding these errors to leading order about the high symmetry configurations $(\theta, \phi) = (0,0)$ and $(\theta, \phi) = (-\pi/4,\pi/4)$, the elastoresistivity coefficients $m_{xx,xx}-m_{xx,yy}$ (in the first configuration) and $2m_{xy,xy}$ (in the second configuration) are misestimated by a factor

\begin{equation}
\label{eq:7}
1 - \bigg[ \delta \theta_{xx}^2 + \delta \theta_{yy}^2 + \delta \phi_{xx}^2 + \delta \phi_{yy}^2 \bigg].
\end{equation}

\noindent The angular alignment errors systematically induce an underestimate of the true elastoresistivity coefficients and come in at second order in the misalignment; even if all angles were off by as much as 5$^{\circ}$ (a typical experimental uncertainty), the total error would only be $\sim 3\%$, and so the high symmetry configurations are relatively insensitive to minor angular offsets.

Additionally, misalignment with respect to the high symmetry configurations also mixes $B_{1g}$ coefficients into a nominal measurement of the $B_{2g}$ symmetry channel and vice versa.  The amount of mixing from the other symmetry channel is proportional to

\begin{equation}
\label{eq:8}
-2 \bigg[ \delta \theta_{xx} \delta \phi_{xx} + \delta \theta_{yy} \delta \phi_{yy} \bigg],
\end{equation}

\noindent which again is at second order in the misalignment.  This mixing due to misalignment could be significant if the relative difference in the magnitudes of the elastoresistivity coefficients in the two symmetry channels is large.  For example, for the specific case of the iron-based superconductors in which $\chi_{_{B_{2g}}}$ diverges, measurement of $m_{xx,xx} - m_{xx,yy}$ is affected by admixture of the much larger $2m_{xy,xy}$ coefficient, whereas measurement of $2m_{xy,xy}$ is essentially unaffected by admixture of a small amount of the much smaller $m_{xx,xx} - m_{xx,yy}$.\cite{kuo_2013}

\hfill \break

\noindent \textbf{(b) \underline{Unequal strain experienced by the two samp-}} \\
\noindent \textbf{\underline{les:}} The differential longitudinal technique relies on both samples experiencing the same homogeneous strain. If the samples experience a different strain due to experimental nonidealities (see next section for a discussion relevant to the specific technique we have employed), this will also affect the deduced elastoresistivity coefficients.

Relative strain errors are of potentially greater concern than misalignment errors since any strain offset error occurs at first order.  Strain offsets also erroneously mix in $A_{1g}$-like coefficients and hence contaminate a nominal nematic susceptibility measurement with the effects of isotropic strain (see discussion in Appendix \ref{appendix:strainmagnitudeerrors}).\cite{footnote1}  Comparison of nominal $B_{1g}$ and $B_{2g}$ coefficients can help bound the amount of $A_{1g}$ contamination (since such rotationally invariant contamination would manifest equally in both $B_{1g}$ and $B_{2g}$ measurements).  Hence, it is still possible to classify which symmetries are broken at the phase transition (i.e., assigning the order parameter to a particular irreducible representation of the space group); however, these concerns motivate development of a technique that does not rely on separate measurements of different samples but which is based instead on measurement of a single sample. This is the primary motivation for adopting either the modified Montgomery technique (Figure~\ref{fig:difflongvstrans} (b)) or the transverse elastoresistance technique (Figure~\ref{fig:difflongvstrans} (c)) that we introduce in Section \ref{sec:transverse}. 

\hfill \break

\noindent \textbf{(c) \underline{Systematic errors originating with the specif-}} \\
\noindent \textbf{\underline{ic technique:}} In addition to the sources of systematic error discussed above, additional errors can be introduced which are specific to the particular technique that is used to strain the samples.  For our experimental realization in which single crystals are glued to the side surface of a PZT stack, these errors are related to differential thermal contraction and strain homogeneity.  We emphasize, however, that strain homogeneity and differential thermal contraction are not necessarily generic to elastoresistivity measurements; alternative methodologies may be able to mitigate or circumvent these particular sources of error while potentially incurring others.

Differential thermal contraction between the sample and the PZT stack on which it is mounted implies that the sample is strained even when no voltage is applied to the piezoelectric. At high temperatures, this is a small effect since the dynamic range over which the sample can be strained exceeds the ``bias'' strain due to such thermal effects.\cite{footnote2}  However, the situation is reversed at cryogenic temperatures.  Upon cooling to 100 K, the crystal experiences a large, anisotropic strain (of about $\sim 0.1 \%$)\cite{hicks_2014} solely from the PZT due to an expansion along its poling direction; depending on the voltage range that is applied to the PZT stack, this can be much larger than the dynamic strain that the PZT can apply due to an applied voltage at this temperature ($\sim 0.01 \%$ at 150 V),\cite{kuo_2013,piezo_data_sheet} which means that the strain experienced by the crystal may not be able to be tuned through zero. So long as the material is still in the regime of linear response, which can be checked, the elastoresistivity coefficients can still be faithfully measured; however, alternative methodologies\cite{hicks_2014} can also be employed which mitigate such effects and more closely yield zero strain conditions.

So far, our analysis has assumed no relaxation of the strain through the thickness of the crystal.  In practice, the strain will relax towards its unstrained edges on a certain length scale determined by the elastic stiffness of the sample and the extent to which the sides of the crystal are ``clamped'' by the epoxy.  To mitigate such effects, samples to be measured must have a thickness that is much smaller than the in-plane dimensions.  Concerns about strain relaxation in the narrow in-plane direction relative to the long in-plane direction can be completely allayed by using a modified Montgomery technique for crystals with a square shape (Figure~\ref{fig:difflongvstrans} (b)).\cite{kuo_2016} Strain transmission has been verified for larger crystals with a similar aspect ratio to those used for the present elastoresistivity measurements by gluing strain gauges directly on the top surface of the crystal.\cite{chu_2012,kuo_2015}

Finally, for crystals with dimensions comparable to the separation distance between two PZT layers, and depending on where the crystal is mounted relative to a PZT layer junction, the strain that the crystal experiences can vary with position on the PZT surface.  The PZT stacks that we have employed have been stacked along their poling direction with an individual layer thickness of $\sim 200$ $\mu$m and a separation between layers of $\sim 50$ $\mu$m.  If a crystal with a width roughly equal to these thicknesses is adhered along the multilayer interface (as is required in a differential longitudinal elastoresistance measurement) and inauspiciously placed in the interfacial separation region, the crystal will experience little strain even with a voltage applied to the piezo.  This sort of inhomogeneity can be ameliorated by spreading the strain transmitting epoxy to encompass more area than just the interface region, attention to placement on the PZT substrate, and use of larger crystals.  Modified techniques can also be readily envisaged that yield a more homogeneous strain.

\section{Transverse Configuration for Probing Nematic Susceptibility in $D_{4h}$}\label{sec:transverse}

\subsection{Ideal Configuration}

\begin{figure}
\centering
\includegraphics[width=\columnwidth]{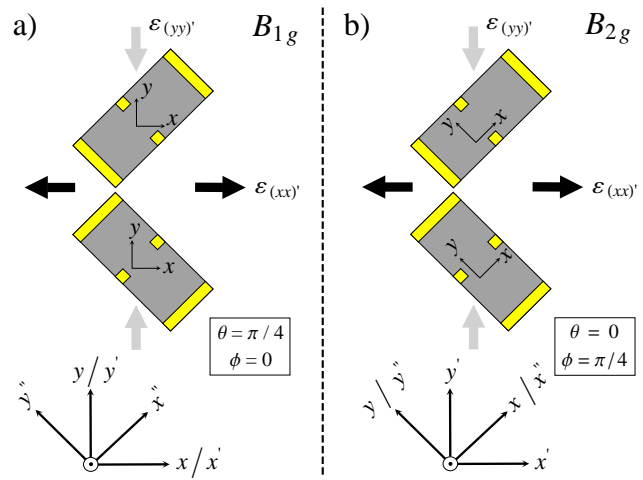}
\caption{Schematic diagrams illustrating transverse elastoresistivity configurations for extracting (a) $m_{xx,xx} - m_{xx,yy}$ (with $(\theta, \phi) = (\pi/4,0)$) and (b) $2m_{xy,xy}$ (right, with $(\theta, \phi) = (0, \pi/4)$).  In (a), one measures the sum of the transverse resistive response to strain $\left(\nicefrac{\Delta \rho}{\rho}\right)_{(xy)''} + \left(\nicefrac{\Delta \rho}{\rho}\right)_{(yx)''}$ to a strain $\epsilon_{(xx)'}-\epsilon_{(yy)'}$ with the transport frame rotated by $\pi/4$ radians relative to the crystal and normal strain frames; the summed elastoresistivity then yields the elastoresistivity coefficients $m_{xx,xx} - m_{xx,yy}$.  In (b), one measures the sum of the transverse resistive response to strain $\left(\nicefrac{\Delta \rho}{\rho}\right)_{(xy)''} + \left(\nicefrac{\Delta \rho}{\rho}\right)_{(yx)''}$ to a strain $\epsilon_{(xx)'}-\epsilon_{(yy)'}$ with the normal strain frame rotated by $\pi/4$ radians relative to the crystal and transport frames; the summed transverse elastoresistivity then yields the elastoresistivity coefficient $2m_{xy,xy}$.}
\label{fig:symmetryconfigb}
\end{figure}

An alternative method for obtaining the same symmetry information involves \textit{transverse} elastoresistance measurements, as depicted in Figure~\ref{fig:symmetryconfigb}.  The underlying intuition is that by appropriately rotating the crystal frame relative to the transport and normal strain frames, the quantity $\left(\nicefrac{\Delta \rho}{\rho}\right)_{(xy)''} + \left(\nicefrac{\Delta \rho}{\rho}\right)_{(yx)''}$ can mix into $\left(\nicefrac{\Delta \rho}{\rho}\right)_{(xx)''} - \left(\nicefrac{\Delta \rho}{\rho}\right)_{(yy)''}$ and hence probe the same $B_{1g}$ and $B_{2g}$ susceptibilities.  Expressing $\left(\nicefrac{\Delta \rho}{\rho}\right)_{(xy)''} + \left(\nicefrac{\Delta \rho}{\rho}\right)_{(yx)''}$ in terms of strains in the normal strain frame and elastoresistivity coefficients in the crystal frame (Appendix \ref{appendix:tensortransform}), we find that

\begin{align}
\label{eq:9}
\left(\nicefrac{\Delta \rho}{\rho}\right)&_{(xy)''}(H_z) + \left(\nicefrac{\Delta \rho}{\rho}\right)_{(yx)''}(H_z) = \\ 
& -\Big[ \epsilon_{(xx)'} - \epsilon_{(yy)'} \Big] \cdot \Big[ 2 m_{xy,xy} \cos(2\theta) \sin(2\phi) \nonumber \\
& \hspace{1cm} + (m_{xx,xx} - m_{xx,yy}) \sin(2\theta) \cos(2\phi) \Big] \nonumber,
\end{align}

\noindent and so the elastoresistivity coefficients corresponding to the $B_{1g}$ and the $B_{2g}$ irreducible representations can be isolated for appropriate high-symmetry configurations.  As depicted in Figure~\ref{fig:symmetryconfigb} (which can be corroborated with \eqref{eq:9}), the appropriate configuration to extract $m_{xx,xx} - m_{xx,yy}$ ($\propto \chi_{_{B_{1g}}}$) via such transverse measurements is to measure the transverse elastoresistivity with currents and transverse voltages directed along the $[110]$ and $[1\bar{1}0]$ crystallographic directions and with strains oriented along the crystalline axes (mathematically, $(\theta, \phi) = (\pi/4,0)$); conversely, extracting $2m_{xy,xy}$ ($\propto \chi_{_{B_{2g}}}$) requires measuring the superposed transverse elastoresistivity with currents and transverse voltages directed along the principal crystalline axes and with strains directed along the $[110]$ and $[1\bar{1}0]$ crystallographic directions (mathematically, $(\theta, \phi) = (0,\pi/4)$).

\begin{figure}
\centering
\includegraphics[width=\columnwidth]{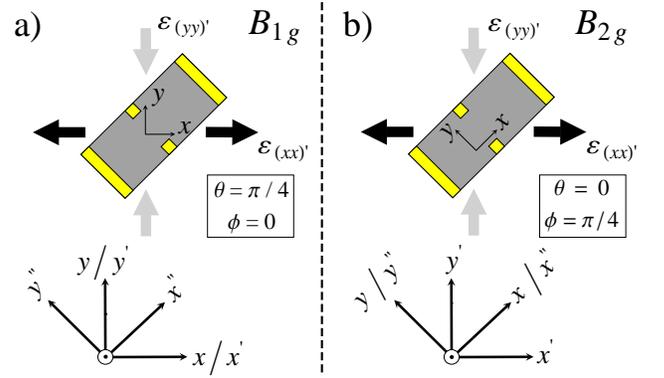}
\caption{Transverse elastoresistivity configurations for extracting (a) $m_{xx,xx} - m_{xx,yy}$ (with $(\theta, \phi) = (\pi/4,0)$) and (b) $2m_{xy,xy}$ (with $(\theta, \phi) = (0, \pi/4)$).  Mounted in these configurations and in the presence of a finite field, the appropriate elastoresistivity coefficients can be measured from $\left(\nicefrac{\Delta \rho}{\rho}\right)_{(xy)''}(H_z) + \left(\nicefrac{\Delta \rho}{\rho}\right)_{(xy)''}(-H_z)$; in zero field, the same coefficients can be extracted with a single measurement $\left(\nicefrac{\Delta \rho}{\rho}\right)_{(xy)''}(H_z=0)$.  The primary advantage of the transverse technique is that one can measure the same elastoresistivity coefficients from a single mounting without $A_{1g}$ contamination.}
\label{fig:symmetryconfigb2}
\end{figure}

The previous discussion was framed in a manner that emphasized the essential similarity between the longitudinal and transverse configurations; however, there is an additional simplification in the transverse configuration that essentially halves the experimentalist's workload.  Since $\left(\nicefrac{\Delta \rho}{\rho}\right)_{(yx)''}(H_z) = \left(\nicefrac{\Delta \rho}{\rho}\right)_{(xy)''}(-H_z)$ from Onsager's relation,\cite{shapiro_2015} one can measure the same elastoresistivity coefficients in the transverse configuration by performing the measurements $\left(\nicefrac{\Delta \rho}{\rho}\right)_{(xy)''}(\pm H_z)$ and taking the sum.  An illustration of this elastoresistivity configuration is given in Figure \ref{fig:symmetryconfigb2} and expressed mathematically as

\begin{align}
\label{eq:10}
\left(\nicefrac{\Delta \rho}{\rho}\right)&_{(xy)''}(H_z) + \left(\nicefrac{\Delta \rho}{\rho}\right)_{(xy)''}(-H_z) = \\ 
& -\Big[ \epsilon_{(xx)'} - \epsilon_{(yy)'} \Big] \cdot \Big[ 2 m_{xy,xy} \cos(2\theta) \sin(2\phi) \nonumber \\
& \hspace{1cm} + (m_{xx,xx} - m_{xx,yy}) \sin(2\theta) \cos(2\phi) \Big] \nonumber.
\end{align}

Instead of dismounting the same crystal and re-mounting in a new configuration, as is required for the differential longitudinal configuration, one need only reverse the orientation of the magnetic field, which is usually simply accomplished \textit{in situ}.  We emphasize that the appropriate elastoresistivity coefficients are given by the \textit{symmetric} combination (i.e., sum) of transverse voltages, in constrast to Hall coefficient measurements, which are given by the anti-symmetric (i.e., difference) combination in magnetic field.  A final comment is that in zero magnetic field, $\left(\nicefrac{\Delta \rho}{\rho}\right)_{(xy)''} = \left(\nicefrac{\Delta \rho}{\rho}\right)_{(yx)''}$, and so only a single measurement is required:

\begin{align}
\label{eq:11}
\left(\nicefrac{\Delta \rho}{\rho}\right)&_{(xy)''}(H_z=0) = \left(\nicefrac{\Delta \rho}{\rho}\right)_{(yx)''}(H_z=0) = \\ 
& -\tfrac{1}{2} \Big[ \epsilon_{(xx)'} - \epsilon_{(yy)'} \Big] \cdot \Big[ 2 m_{xy,xy} \cos(2\theta) \sin(2\phi) \nonumber \\
& \hspace{1cm} + (m_{xx,xx} - m_{xx,yy}) \sin(2\theta) \cos(2\phi) \Big] \nonumber.
\end{align}

\subsection{Sources of Error}

As for the differential longitudinal configuration, angular misalignment and strain magnitude errors can also manifest in the transverse method.  In addition, misalignment of the contacts used to measure the transverse voltages can lead to some amount of longitudinal resistivity $\rho_{xx}$ seeping into an intended measurement of $\rho_{xy}$, in which case it is necessary to determine an appropriate method to correctly subtract such longitudinal contamination.  In this section, we address each of these concerns, quoting a few main results (whose full derivation appears in Appendices \ref{appendix:alignmenterrors}, \ref{appendix:strainmagnitudeerrors}, and \ref{appendix:transverseisolation}) and emphasizing the advantages of the transverse setup.  In particular we note the principal advantage of the transverse technique is that the measurement does not suffer from $A_{1g}$ contamination.

\hfill \break

\noindent \textbf{(a) \underline{Angular misalignment:}} Alignment errors propagate in analogous ways as for the differential longitudinal case (Section \ref{sec:difflongerrors}).  Assuming that we intended to measure $\left(\nicefrac{\Delta \rho}{\rho}\right)_{(xy)''}(H_z)$ and $\left(\nicefrac{\Delta \rho}{\rho}\right)_{(yx)''}(H_z)$ in a $(\theta, \phi)$ configuration but actually mounted at $(\theta + \delta \theta_{xy}, \phi + \delta \phi_{xy})$ and $(\theta + \delta \theta_{yx}, \phi + \delta \phi_{yx})$ (respectively), we obtain the analog of \eqref{eq:7} for the propagated error and \eqref{eq:8} for the contamination from the other symmetry channel (one need only interchange subscripts $xx \leftrightarrow xy$ and $yy \leftrightarrow yx$; see Appendix \ref{appendix:alignmenterrors}); again, in the high symmetry configurations, the leading errors are at second order in the angular misalignment and consequently lead to negligibly small systematic errors.

\hfill \break

\noindent \textbf{(b) \underline{Unequal strain experienced by the two samp-}} \\
\noindent \textbf{\underline{les:}} Strain offset errors are fundamentally different than in the differential longitudinal configuration, which is the primary advantage of the transverse geometry. In zero magnetic field, the quantity $\left(\nicefrac{\Delta \rho}{\rho}\right)_{(xy)''}$ is fundamentally immune to strain offset errors (indeed, there is no offset since only a single sample is needed, in contrast to the differential  technique).  Furthermore, the isotropic strains that are experienced by the crystal cannot generate a transverse voltage: that is, rotationally invariant ($A_{1g}$) strains cannot produce directionally oriented ($B_{1g}$ or $B_{2g}$) resistivity changes (see Appendix \eqref{eq:c6}, with $m_{xy,xx} = m_{xy,zz} = 0$ in vanishing field).  In a finite field, a second measurement is needed, but since Onsager gives $\left(\nicefrac{\Delta \rho}{\rho}\right)_{(yx)''}(H_z) = \left(\nicefrac{\Delta \rho}{\rho}\right)_{(xy)''}(- H_z)$, one can measure the induced resistivity changes without re-gluing the crystal; therefore, one can be sure that the strain offset errors in the $\left(\nicefrac{\Delta \rho}{\rho}\right)_{(xy)''}(\pm H_z)$ measurements are exactly zero.  \textit{This is the primary advantage of the transverse technique.}

\hfill \break

\begin{figure}
\centering
\includegraphics[width=\columnwidth]{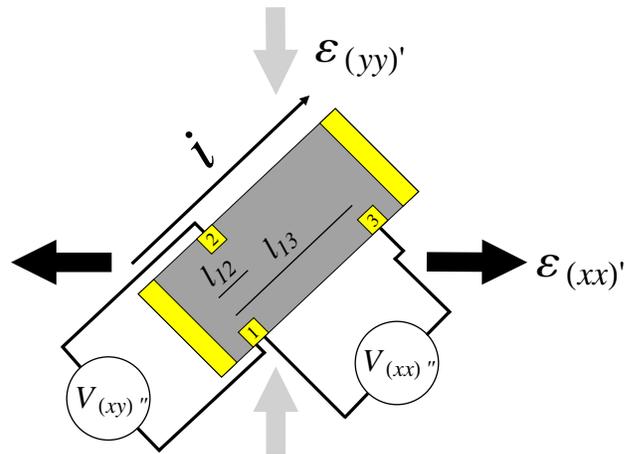}
\caption{Schematic diagram showing contact geometry for a practical transverse elastoresistance measurement.  Contacts 1 and 2 are used to measure $R_{(xy)''}$; however, partial misalignment can lead to $R_{(xx)''}$ contamination in a nominal $R_{(xy)''}$ measurement.  This contamination can be accounted for by subtracting out a down-weighted longitudinal contribution, with the down-weighting given by the factor $\Delta_{\ell}$.  $\Delta_{\ell}$ is defined as the ratio of the misalignment distance $l_{12}$ to the distance between longitudinal contacts $l_{13}$, which is related to the ratio of the transverse and longitudinal voltages on a free-standing crystal: $\Delta_{\ell} \equiv \nicefrac{l_{12}}{l_{13}} = \nicefrac{V_{(xy)''}(H=0,\textrm{free-standing})}{V_{(xx)''}(H=0,\textrm{free-standing})}$ (see Appendix \ref{appendix:transverseisolation}).  This subtraction procedure is analogous to the anti-symmetrization procedure that is used for Hall effect measurements.  To use this geometry to probe $\chi_{_{B_{1g}}}$ or $\chi_{_{B_{2g}}}$ requires mounting the crystal such that $(\theta,\phi) = (\pi/4,0)$ or $(\theta,\phi) = (0,\pi/4)$, respectively, as described in the main text.}
\label{fig:transelastocontacts}
\end{figure}

\noindent \textbf{(c) \underline{Subtracting $\rho_{xx}$ contamination from a transv-}} \\
\noindent \textbf{\underline{erse measurement:}} One additional complication in measuring $\left(\nicefrac{\Delta \rho}{\rho}\right)_{(xy)''}(H_z)$, however, is subtracting out any unwanted contributions from $\rho_{xx}$ in a putative $\rho_{xy}$ measurement due to unintentional contact misalignment.  In a typical Hall measurement of $\rho_{xy}$, one can use the fact that the transverse force on the electrons is odd in the magnetic field and hence anti-symmetrize the data in field to subtract out $\rho_{xx}$.  This approach does not work for transverse elastoresistivity, where the symmetry-motivated elastoresistivity coefficients of interest are themselves even in the magnetic field (despite coming from a measurement of $\left(\nicefrac{\Delta \rho}{\rho}\right)_{(xy)''} (H_z)$).  Instead, one needs to characterize the amount of longitudinal $\rho_{xx}$ contamination in terms of the geometry of the electrical contacts themselves.  Parametrizing this geometrical misalignment by a parameter $\Delta_{\ell}$, one accounts for such contamination by simultaneously measuring the longitudinal elastoresistance $\left(\nicefrac{\Delta \rho}{\rho}\right)_{(xx)''} (H_z)$ from a second pair of contacts (Figure \ref{fig:transelastocontacts}) and precisely subtracting out the down-weighted contribution $\Delta_{\ell} \left[ \left(\nicefrac{\Delta \rho}{\rho}\right)_{(xx)''} (H_z) \right]$ from $\left(\nicefrac{\Delta \rho}{\rho}\right)_{(xy)''} (H_z)$.  The full subtraction procedure is derived and outlined in Appendix \ref{appendix:transverseisolation}.

\section{Transverse Elastoresistivity Measurements of BaFe$_2$As$_2$}

In order to demonstrate the efficacy of the new transverse  configuration, we chose to measure the $2m_{xy,xy}$ elastoresistivity coefficient of the representative iron-pnictide BaFe$_2$As$_2$.  Since we have already extracted this coefficient using the differential longitudinal method,\cite{kuo_2015} this allows for a direct comparison between the two configurations.  As we demonstrate below, values of $2m_{xy,xy}$ extracted from the two techniques agree in their temperature dependence, revealing a nematic instability in the $B_{2g}$ symmetry channel.

\subsection{Experimental Methods}

Single crystals of BaFe$_2$As$_2$ were grown from a self-flux method as described elsewhere.\cite{sefat_2008,chu_2009}  The crystals grow as thin plates, with the $c$-axis perpendicular to the plane of the plates and natural facets along the in-plane principal tetragonal axes.   A representative, as-grown, rectangular ($1.6 \textrm{ mm} \times 0.67 \textrm{ mm} \times 0.029 \textrm{ mm}$) crystal was selected for the transport measurements.  X-ray diffraction was used to confirm that the crystallographic [100] and [010] axes were oriented along the length/width of the sample.  Electrical contacts (with current sourced along the $[100]$ tetragonal direction) were affixed to gold-sputtered pads with Dupont 4929N silver paste.

\begin{figure}
\centering
\includegraphics[width=0.9\columnwidth]{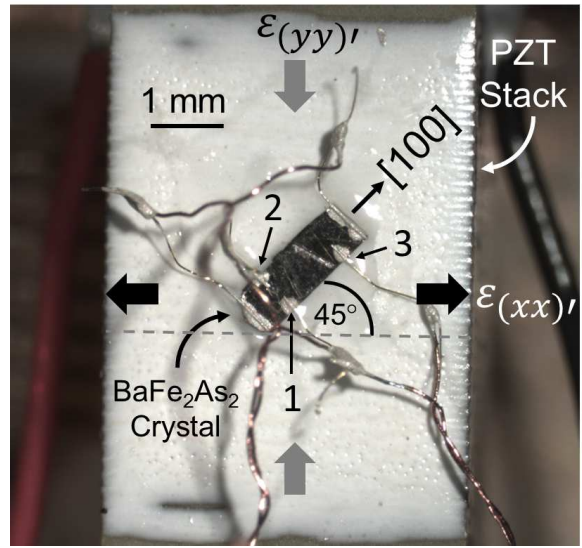}
\caption{Photograph showing a BaFe$_2$As$_2$ crystal affixed to the surface of a PZT stack and mounted in the transverse elastoresistivity configuration $(\theta,\phi) = (0,\pi/4)$, appropriate for measuring $2m_{xy,xy}$ ($\propto \chi_{_{B_{2g}}}$).  Contacts 1, 2, and 3 are labeled with reference to Figure \ref{fig:transelastocontacts}. The twisted pairs used for performing the voltage measurements are evident.}
\label{fig:gluedsample}
\end{figure}

Prior to gluing the sample to the PZT stack, the temperature dependence of the resistances $R_{(xx)''}$ and $R_{(xy)''}$ were measured for the free-standing, unstrained crystal in order to pre-characterize the contact geometry.  The unstrained $R_{(xx)''}$ is also used for normalizing the elastoresistance data.  The samples were then glued to the top surface of a PZT piezoelectric stack (Part Number PSt 150/5$\times$5/7 cryo 1, from Piezomechanik GmbH) using ITW Devcon five minute epoxy spread uniformly across the bottom of the crystal (Figure \ref{fig:gluedsample}).  The orientation of the crystal axes of the sample with respect to the principal axes of the PZT stack was initially determined by eye, such that the long axis of the transport bar was at an angle $\phi$ of approximately 45$^{\circ}$ with respect to the PZT stack. The angle was subsequently determined more precisely from measurements of the photograph shown in Figure \ref{fig:gluedsample} to be $45.4^{\circ} \pm 0.2^{\circ}$.  Mutually transverse strain gauges (Part Number WK-05-062TT-350, from Vishay Precision Group) were glued to the other side of the PZT stack in order to measure the strains $\epsilon_{(xx)'}$ and $\epsilon_{(yy)'}$ \textit{in situ}.

The PZT stack was mounted on the coldhead of a specially adapted probe and cooled in exchange gas in a Janis flow cryostat.  Temperature was controlled with a Lake Shore 340 temperature controller, with a stability of $\pm 50$ mK.\cite{footnote3}  The resistances of both the sample and the strain gauges were measured using Stanford Research Systems SR830 lock-in amplifiers; for the sample, Stanford Research Systems Model SR560 preamplifiers were also used.  AC excitation currents of 1 mA and 0.1 mA were used for the sample and strain gauges, respectively.

Elastoresistance data at a fixed temperature were acquired from changes in the resistances of the sample ($R_{(xx)''}$ and $R_{(xy)''}$) and of both strain gauges while sweeping the voltage applied to the PZT between $-50$ V and $+150$ V.  The voltage was swept stepwise in 4 V increments with a delay of 0.25 s between steps; the measured elastoresistance did not depend on this sweep rate scheme, nor were there any observed heating effects.\cite{riggs_2015}  Three full voltage loops were taken for each temperature setpoint; after completing these loops, the temperature was then stepped to a new setpoint and allowed to stabilize before performing the next elastoresistance measurement.

\subsection{Results}

\begin{figure}
\centering
\includegraphics[width=\columnwidth]{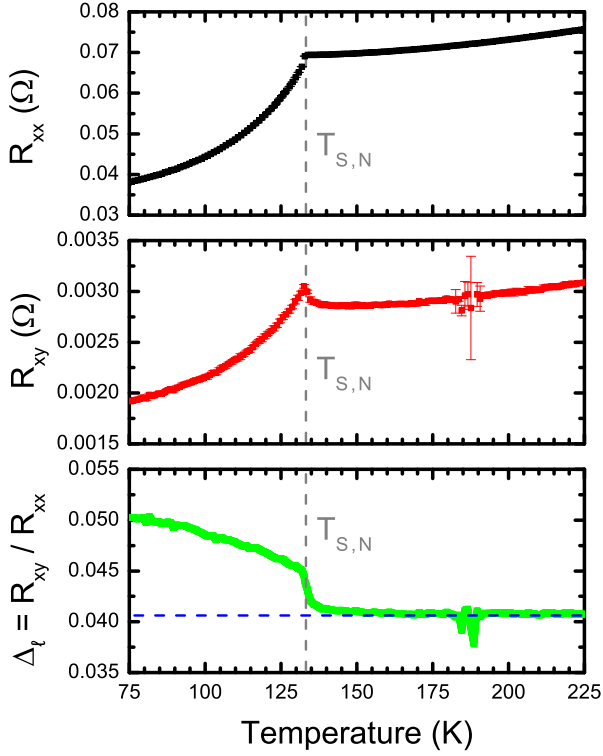}
\caption{Resistance measurements for the BaFe$_2$As$_2$ sample showing $R_{xx}$ (top panel), $R_{xy}$ (middle panel), and the ratio $\Delta_{\ell} \equiv \nicefrac{R_{xy}}{R_{xx}}$ (bottom panel). For temperatures above the coupled structural and magnetic transition at $T_{S,N}$ (vertical dashed gray line), $\Delta_{\ell}$ is small and temperature-independent; by taking the average value for temperatures $> 150$ K, we estimate $\Delta_{\ell} \sim 4.1 \pm 0.1 \%$ (horizontal dashed blue line).  Since this measurement was performed on a free-standing (unglued) sample in zero magnetic field, the measured $R_{xy}$ in the tetragonal state is due to $R_{xx}$ contamination from contact misalignment; $\Delta_{\ell}$ characterizes the physical extent of this misalignment.}
\label{fig:deltaell}
\end{figure}

The temperature dependence of the longitudinal and transverse resistances for the free-standing (unstrained) sample are shown in Figure \ref{fig:deltaell}. The longitudinal resistance $R_{xx}$ follows the usual temperature dependence for this material, exhibiting  a downturn at the coupled structural-magnetic phase transition at $T_{S,N}$ = 134 K. A finite $R_{xy}$ is measured even for the unstrained sample (middle panel of Figure \ref{fig:deltaell}) due to misalignment of the contacts used for the transverse voltage measurement. As can be seen, for temperatures above $T_{S,N}$, the ratio $R_{xy}$/$R_{xx}$ = $\Delta_{\ell}$ is temperature-independent, with a value of $\sim 0.041 \pm 0.01$. A small deviation from this constant value can be discerned for temperatures just above $T_{S,N}$, presumably due to residual strains in the sample.\cite{man_2015} The subsequent discussion and analysis refers solely to temperatures above $T_{S,N}$, for which the material is tetragonal; below this temperature, the crystal structure is orthorhombic, and the transverse resistance reflects an admixture of effects arising from longitudinal contamination, electronic anisotropy associated with the orthorhombicity, twin domain populations, and twin boundary motion. 

\begin{figure}
\centering
\includegraphics[width=\columnwidth]{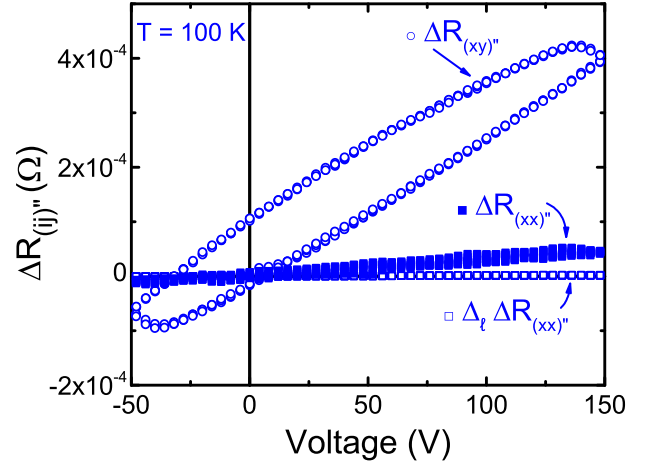}
\caption{Representative data for BaFe$_2$As$_2$ showing strain-induced changes in resistance $\Delta R_{(xy)''}$, $\Delta R_{(xx)''}$, and $\Delta_{\ell} \Delta R_{(xx)''}$ at a temperature of 100 K as a function of the voltage applied to the PZT stack.  Values of $\Delta_{\ell}$ are extracted from Figure~\ref{fig:deltaell} as described in the main text.  In order to correct for $\rho_{xx}$ contamination, the relatively small amount $\Delta_{\ell} \Delta R_{(xx)''}$ is subtracted from $\Delta R_{(xy)''}$.}
\label{fig:100Kelastoresistance}
\end{figure}

Representative strain-induced resistance changes $\Delta R_{(xy)''}$, $\Delta R_{(xx)''}$, and $\Delta_{\ell} \Delta R_{(xx)''}$ for three voltage sweeps at 100 K are shown in Figure \ref{fig:100Kelastoresistance}.  The elastoresistive response in $\Delta R_{(xy)''}$ is significantly larger than $\Delta R_{(xx)''}$, by a factor of $\sim 9$ at 150 V.  To subtract out the $\rho_{xx}$ contamination, one down-weights $\Delta R_{(xx)''}$ even further by $\Delta_{\ell}$; the resulting correction ($\sim 0.4 \%$) is essentially negligible on the scale of the $\Delta R_{(xy)''}$ response.  The hysteretic behavior evident in $\Delta R_{(xy)''}$ is an intrinsic property of the PZT stack.

\begin{figure}
\hspace{-0.6cm} \centering
\includegraphics[width=1.05\columnwidth]{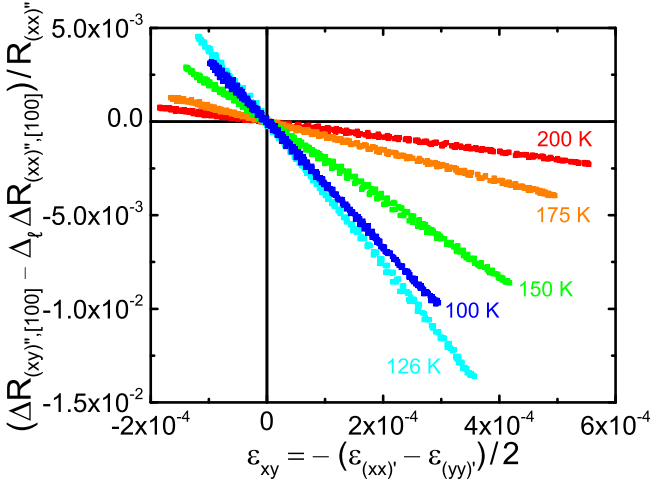}
\caption{Representative data showing the temperature dependence of the elastoresistive response of BaFe$_2$As$_2$ in the transverse configuration $(\theta,\phi) = (0,\pi/4)$ as a function of the induced shear strain ($\epsilon_{xy} = -\frac{1}{2}(\epsilon_{(xx)'} - \epsilon_{(yy)'})$) experienced by the crystal.  This is the appropriate configuration for measuring $2m_{xy,xy}$ ($\propto \chi_{_{B_{2g}}}$).  Slopes have been corrected for $\rho_{xx}$ contamination, as described in the main text and Appendix \ref{appendix:transverseisolation}.  All responses were linear in the applied strain for all measured temperatures.}
\label{fig:slopetempdep}
\end{figure}

\begin{figure}
\centering
\includegraphics[width=\columnwidth]{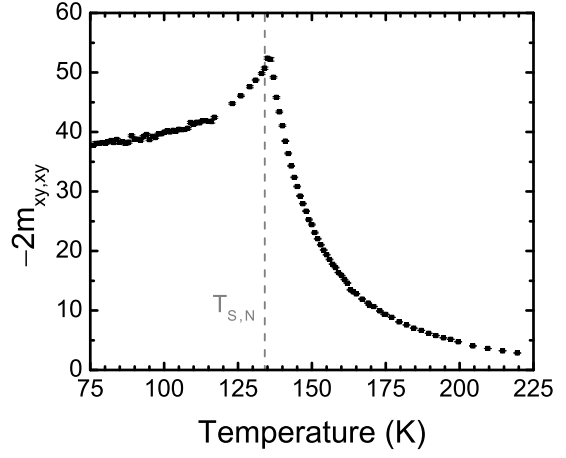}
\caption{Temperature dependence of the elastoresistivity coefficient $2m_{xy,xy}$ of BaFe$_2$As$_2$ from the transverse configuration. Error bars represent 95\% confidence intervals from the linear fits to the elastoresistive slopes at each temperature.  The vertical dashed bar marks the coincident structural and magnetic transition temperature $T_{S,N}$ of the sample.}
\label{fig:2mxyxytempdep}
\end{figure}

Representative transverse elastoresistance data, which have been corrected for the small longitudinal contamination as described above, are shown in Figure \ref{fig:slopetempdep} as a function of applied shear strain ($\epsilon_{xy} = -\frac{1}{2}(\epsilon_{(xx)'} - \epsilon_{(yy)'})$) for a variety of temperatures.  The applied shear strain is relative to the strain experienced by the crystal with 0 V applied to the PZT stack.  Because of the combined effects of gluing to the PZT stack and differential thermal contraction between the crystal and PZT, the applied shear strain is \textit{not} relative to the zero strain state of the crystal.  As can be seen, all measured elastoresistances are linear in (relative) shear strain for all measured temperatures.  The elastoresistivity coefficient $2m_{xy,xy}$ at each temperature is extracted from the slopes in Figure \ref{fig:slopetempdep} and a multiplicative factor of the crystal's length to its width used to convert the resistance ratio to a resistivity ratio ($\left( \nicefrac{\Delta \rho}{\rho} \right)_{(xy)''} = \frac{l}{w} \left( \nicefrac{\Delta R_{(xy)''}}{R_{xx}} \right)$; see Appendix \ref{appendix:transverseisolation}).  The temperature dependence of the resulting values of $2m_{xy,xy}$ is shown in Figure \ref{fig:2mxyxytempdep}.  As has been previously shown,\cite{kuo_2015,kuo_2016,kuo_thesis} $2m_{xy,xy}$ progressively increases on cooling, reaches its peak at the coupled structural and magnetic transition temperature, and then gradually decreases on further cooling.  The maximum value of $2m_{xy,xy}$ is $\sim 52$, much larger than that of a typical metal ($\sim 1$).

\begin{table*}
\centering
\caption{\textbf{Fit Parameters from Transverse, Differential Longitudinal, and Modified Montgomery Methods}}
Fit to $2m_{xy,xy} = \frac{\lambda}{a_0(T-\theta)} + 2m_{xy,xy}^{0}$; uncertainties represent 95\% confidence intervals from a least squares fitting routine
\makebox[\textwidth][c]{
\begin{tabular}{c | c | c | c}
\hline\hline
Parameter & Transverse Method & Differential Longitudinal Method\cite{kuo_thesis} & Modified Montgomery Method\cite{kuo_2016}\Tstrut \\
\hline \hline
$2m_{xy,xy}^{0}$ & $6.7 \pm 0.5$ & $4.7 \pm 0.98$ & $7.7 \pm 0.3$ \rule{0pt}{1.5\normalbaselineskip} \\[2ex]
\hline
$\nicefrac{\lambda}{a_0}$ (K) & $-909 \pm 16$ & $-897 \pm 84$ & $-1540 \pm 13$ \rule{0pt}{1.5\normalbaselineskip} \\[2ex]
\hline
$\theta$ (K) & $120 \pm 0.9$ & $124.9 \pm 1.6$ & $109 \pm 0.7$ \rule{0pt}{1.5\normalbaselineskip} \\[2ex]
\hline
\end{tabular}}
\label{table1}
\end{table*}

In accordance with our previous analysis of differential longitudinal elastoresistance measurements,\cite{chu_2012,kuo_2013} the diverging temperature dependence of $2m_{xy,xy}$ can be fit well to the Curie-Weiss form

\begin{equation}
\label{eq:12}
2m_{xy,xy} = \frac{\lambda}{a_0(T-\theta)} + 2m_{xy,xy}^0.
\end{equation}

\begin{figure}
\centering
\includegraphics[width=\columnwidth]{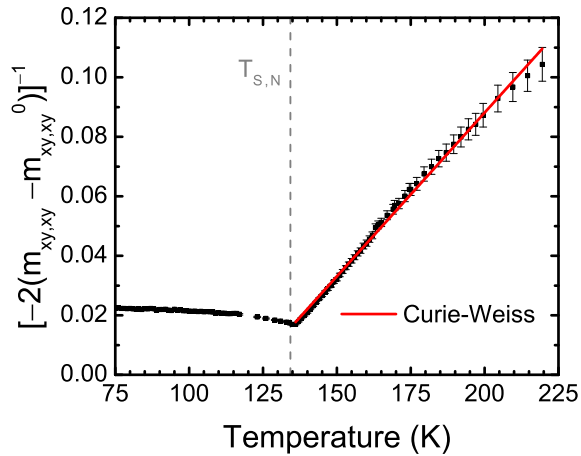}
\caption{Temperature dependence of $[-2(m_{xy,xy} - m_{xy,xy}^0)]^{-1}$, proportional to the inverse nematic susceptibility $\chi_{_{B_{2g}}}^{-1}$ in this configuration.  Error bars represent 95\% confidence intervals from both the linear fits to the elastoresistive slopes at each temperature and the estimation of $2m_{xy,xy}^0$.  The red line shows a linear fit (i.e., Curie-Weiss functional form) between $136$ K and $220$ K, with fit parameters given in Table \ref{table1}.  The vertical dashed line marks the coincident structural and magnetic transition temperature $T_{S,N}$ of the sample ($T_{S,N} = 134$ K).}
\label{fig:2mxyxyCWfit}
\end{figure}

\noindent Directly fitting the temperature dependence of $2m_{xy,xy}$ to the Curie-Weiss form \eqref{eq:12}, the temperature-independent fit parameter $2m_{xy,xy}^{0}$ can be estimated and then used to plot the temperature dependence of the inverse susceptibility $[-2(m_{xy,xy} - m_{xy,xy}^0)]^{-1}$ (which, for an exact Curie-Weiss form, is linear in temperature).  Fitting over the temperature range of 136 K to 220 K results in an estimate of $2m_{xy,xy}^{0} = 6.7 \pm 0.5$, which we use to plot the temperature dependence of $[-2(m_{xy,xy} - m_{xy,xy}^0)]^{-1}$ as in Figure \ref{fig:2mxyxyCWfit}.  Further details on the evaluation of the goodness of fit and on the temperature window used are discussed in Appendix \ref{appendix:cwparameterfit}.  The slope and intercept of $[-2(m_{xy,xy} - m_{xy,xy}^0)]^{-1}$ yield estimates of $\nicefrac{\lambda}{a_0} = -909 \pm 16$ K and $\theta = 120 \pm 0.9$ K, as given in Table \ref{table1}.  As we have previously discussed, the observation of such a Curie-Weiss susceptibility with a Weiss temperature $\theta$ close to the coupled structural and magnetic transition definitively establishes the ferroelastic phase transition in BaFe$_2$As$_2$ to be pseudo-proper (i.e., strain is not the primary order parameter of the transition but does have the same symmetry as the order parameter).\cite{kuo_2013,chu_2012}  The physical origin of the electronic nematic order that drives this phase transition remains a subject of ongoing research (for example, see Fernandes \textit{et al.}\cite{fernandes_2014} for a recent review and discussion).

The estimated Curie-Weiss fit parameters from the transverse method can also be compared to the parameter estimates from earlier measurements of $2m_{xy,xy}$ by the differential longitudinal\cite{kuo_thesis} and modified Montgomery methods\cite{kuo_2016} (see Table \ref{table1}).  All three measurements agree in their divergent temperature dependence, which evinces the existence and onset of a nematic order parameter.  The differential longitudinal and transverse measurements agree within $\sim 4.0 \%$ in their estimate of $\theta$ (which characterizes a bare mean field nematic critical temperature) and agree within $\sim 1.3 \%$ in their estimates of $\nicefrac{\lambda}{a_0}$; meanwhile, there is a larger discrepancy between the estimates of $\theta$ and $\nicefrac{\lambda}{a_0}$ as obtained from the modified Montgomery and transverse methods.

The quantitative variations in the estimated fit parameters between the three methods presumably reflect systematic differences in the physical environment in which the three experiments are performed.  Strictly, elastoresistivity coefficients are defined in the limit of vanishing strain; however, this limit is not precisely realized in any of the three methods.  The elastoresistivity coefficients as extracted from the differential longitudinal and transverse methods are measured relative to the strain state of the crystal with 0 V applied to the PZT (i.e., relative to a state with some residual built-in isotropic and anisotropic strain due to adhesion to the strain-transmitting substrate and differential thermal contraction), while the elastoresistivity coefficients as extracted from the modified Montgomery method are measured relative to a ``$B_{2g}$ neutral point''\cite{kuo_2016} where the anisotropic strain is tuned to zero by applying a finite voltage to the PZT (until the longitudinal resistivities $\rho_{xx}$, $\rho_{yy}$ are equal) but where the isotropic strain is explicitly nonzero. Furthermore, since the physical dimensions of the crystals vary between the three studies (``matchstick'' rectangular bars for the differential longitudinal measurements; square or rectangular plates with a nearly 2:1 aspect ratio for the modified Montgomery and transverse methods), the effect of strain relaxation due to the geometry of the crystals could plausibly contribute to systematic variation in the fit parameter estimates as well. The exact reasons for the quantitative differences, however, are not yet perfectly understood.

\section{Conclusion}

In writing this paper, we have had two overarching goals.  First, building on the elastoresistivity formalism that we have introduced in recent publications,\cite{kuo_2013,shapiro_2015} we have proposed an alternative method to probe the $B_{1g}$ and $B_{2g}$ components of the elastoresistivity tensor for a tetragonal material via transverse elastoresistivity measurements.  We have quantified the effects of various experimental nonidealities that affect both the earlier differential longitudinal elastoresistance methods and the new transverse elastoresistance method and have shown that the transverse scheme has certain specific advantages.  In particular, the transverse technique enables measurement of $m_{xx,xx} - m_{xx,yy}$ or $2m_{xy,xy}$ via a single measurement. Importantly, since isotropic strains cannot induce a transverse elastoresistivity $\left( \nicefrac{\Delta \rho}{\rho} \right)_{(xy)''}$, and since $2m_{xy,xy}$ was extracted from a single measurement, the new method is fundamentally immune to $A_{1g}$-like strains and strain offset errors.

Second, we have used the representative iron-pnictide BaFe$_2$As$_2$ to explicitly demonstrate the viability of the transverse elastoresistivity configuration.  The new method corroborates the earlier finding of a Curie-Weiss-like $2m_{xy,xy}$ elastoresistivity coefficient in this material, signaling the divergent nematic susceptibility in the $B_{2g}$ symmetry channel.  To our knowledge, this is the first discussion and measurement of transverse elastoresistance for any material.

\begin{acknowledgments}

The authors thank P. Walmsley and Eric M. Spanton for useful discussions.  Work at Stanford was supported by the U.S. DOE, Office of Basic Energy Sciences, under contract DE-AC02-76SF00515.

\end{acknowledgments}

\onecolumngrid
\newpage
\appendix

\section{Elastoresistivity Tensor for $D_{4h}$ and Transformation Properties}\label{appendix:tensortransform}

The explicit form of the elastoresistivity tensor for $D_{4h}$ point group symmetry is\cite{shapiro_2015}

\begin{equation}
\label{eq:a1}
m^{\textrm{tetragonal}}_{ij,kl}(H_z) =
\begingroup
\renewcommand*{\arraystretch}{1.25}
\begin{pmatrix}
m_{xx,xx} & m_{xx,yy} & m_{xx,zz} & 0 & 0 & 0 & 0 & 0 & 0 \\
m_{xx,yy} & m_{xx,xx} & m_{xx,zz} & 0 & 0 & 0 & 0 & 0 & 0 \\
m_{zz,xx} & m_{zz,xx} & m_{zz,zz} & 0 & 0 & 0 & 0 & 0 & 0 \\
\hhline{~~~~~--~~}
0 & 0 & 0 & m_{yz,yz} & m_{yz,yz} & \multicolumn{1}{|c}{m_{yz,zx}} & \multicolumn{1}{c|}{m_{yz,zx}} & 0 & 0 \\
0 & 0 & 0 & m_{yz,yz} & m_{yz,yz} & \multicolumn{1}{|c}{-m_{yz,zx}} & \multicolumn{1}{c|}{-m_{yz,zx}} & 0 & 0 \\
\hhline{~~~----~~}
0 & 0 & 0 & \multicolumn{1}{|c}{m_{yz,zx}} & \multicolumn{1}{c|}{m_{yz,zx}} & m_{yz,yz} & m_{yz,yz} & 0 & 0 \\
0 & 0 & 0 & \multicolumn{1}{|c}{-m_{yz,zx}} & \multicolumn{1}{c|}{-m_{yz,zx}} & m_{yz,yz} & m_{yz,yz} & 0 & 0 \\
\hhline{-----~~~~}
\multicolumn{1}{|c}{m_{xy,xx}} & m_{xy,xx} & \multicolumn{1}{c|}{m_{xy,zz}} & 0 & 0 & 0 & 0 & m_{xy,xy} & m_{xy,xy} \\
\multicolumn{1}{|c}{-m_{xy,xx}} & -m_{xy,xx} & \multicolumn{1}{c|}{-m_{xy,zz}} & 0 & 0 & 0 & 0 & m_{xy,xy} & m_{xy,xy}\\
\hhline{---~~~~~~}
\end{pmatrix}.
\endgroup
\end{equation}

\noindent This tensor has 10 independent coefficients, all implicitly dependent on the magnetic field. Those coefficients (of which there are seven) that have an even number of $x$ or an even number of $y$ indices are correspondingly even functions of the magnetic field due to the $\sigma_x$ and $\sigma_y$ symmetry constraints (where $\sigma_x$ and $\sigma_y$ are mirror operations about the $yz$ and $xz$ planes, respecively).  Conversely, those coefficients (of which there are three, demarcated by surrounding boxes) that have an odd number of $x$ or an odd number of $y$ indices are odd functions of the magnetic field (and hence vanish in zero magnetic field).  The symmetry properties of this tensor are described in detail elsewhere.\cite{shapiro_2015}

In a given elastoresistivity measurement, what one measures is $(\nicefrac{\Delta \rho}{\rho})_{(ij)''}$ (i.e., the normalized resistivity change in the transport frame) to an applied strain $\epsilon_{(kl)'}$ in the strain frame, and what one seeks to extract are terms in the elastoresistivity tensor $m_{ij,kl}$ in the crystal frame.  These quantities are related by suitable transformation matrices $\hat{\alpha}_{\theta}, \hat{\alpha}_{\phi}$ according to \eqref{eq:5} of the main text with the $\hat{\alpha}_{\phi}$, $\hat{\alpha}_{\theta}$ given by rotational transformations of the form

\begin{equation}
\label{eq:a2}
\hat{\alpha}_{\phi} =
\begin{pmatrix}
\cos^2(\phi) & \sin^2(\phi) & 0 & 0 & 0 & 0 & 0 & \cos(\phi)\sin(\phi) & \cos(\phi)\sin(\phi) \\
\sin^2(\phi) & \cos^2(\phi) & 0 & 0 & 0 & 0 & 0 & -\cos(\phi)\sin(\phi) & -\cos(\phi)\sin(\phi) \\
0 & 0 & 1 & 0 & 0 & 0 & 0 & 0 & 0 \\
0 & 0 & 0 & \cos(\phi) & 0 & 0 & -\sin(\phi) & 0 & 0 \\
0 & 0 & 0 & 0 & \cos(\phi) & -\sin(\phi) & 0 & 0 & 0 \\
0 & 0 & 0 & 0 & \sin(\phi) & \cos(\phi) & 0 & 0 & 0 \\
0 & 0 & 0 & \sin(\phi) & 0 & 0 & \cos(\phi) & 0 & 0 \\
-\cos(\phi)\sin(\phi) & \cos(\phi)\sin(\phi) & 0 & 0 & 0 & 0 & 0 & \cos^2(\phi) & -\sin^2(\phi) \\
-\cos(\phi)\sin(\phi) & \cos(\phi)\sin(\phi) & 0 & 0 & 0 & 0 & 0 & -\sin^2(\phi) & \cos^2(\phi)
\end{pmatrix}
\end{equation}

\noindent and analogously for $\hat{\alpha}_{\theta}$.

Prior to discussing the implications of these transformations for the elastoresistivity tensor, it is elucidating to focus first on the effect of these rotational transformations on the strain and normalized change in resistivity tensors individually.  If the crystal frame is oriented relative to the strain frame by an angle $\phi$, then the strains experienced by the crystal are related to the shearless strains in the normal strain frame by $\epsilon_{kl} = \hat{\alpha}_{\phi} \epsilon_{(kl)'}$; explicitly, this gives

\begin{equation}
\label{eq:a3}
\begin{pmatrix} \epsilon_{xx} \\ \epsilon_{yy} \\ \epsilon_{zz} \\ \epsilon_{yz} \\ \epsilon_{zy} \\ \epsilon_{zx} \\ \epsilon_{xz} \\ \epsilon_{xy} \\ \epsilon_{yx} \end{pmatrix} = \begin{pmatrix} \epsilon_{(xx)'} \cos^2\phi + \epsilon_{(yy)'} \sin^2\phi \\ \epsilon_{(xx)'} \sin^2\phi + \epsilon_{(yy)'} \cos^2\phi \\ \epsilon_{(zz)'} \\ 0 \\ 0 \\ 0 \\ 0 \\ -\frac{1}{2}(\epsilon_{(xx)'} - \epsilon_{(yy)'})\sin(2\phi) \\[3pt] -\frac{1}{2}(\epsilon_{(xx)'} - \epsilon_{(yy)'})\sin(2\phi) \end{pmatrix}.
\end{equation}

\noindent By rotating the crystal relative to the normal strain frame, the shearless strains in the normal strain frame are experienced as both normal and shear strains in the crystal frame, with the amount of shear characterized by $\sin(2\phi)$ according to $\epsilon_{xy} = \epsilon_{yx} = -\frac{1}{2}(\epsilon_{(xx)'} - \epsilon_{(yy)'})\sin(2\phi)$.

Exactly analogous relations exist for transport measurements expressed in the transport and crystal frames.  Assuming that the two frames are rotated relative to each other by an angle $\theta$, then what one measures in the transport frame is related to the normalized change in resistivity in the crystal frame by $\left(\nicefrac{\Delta \rho}{\rho}\right)_{(ij)''} = \hat{\alpha}_{\theta} \left(\nicefrac{\Delta \rho}{\rho}\right)_{ij}$.  Working this out explicitly for in-plane transport measurements,

\begin{subequations}
\label{eq:a4}
\begin{align}
\left(\nicefrac{\Delta \rho}{\rho}\right)_{(xx)''}(H_z) &= \left(\nicefrac{\Delta \rho}{\rho}\right)_{xx}\cos^2\theta + \left(\nicefrac{\Delta \rho}{\rho}\right)_{yy}\sin^2\theta + \frac{1}{2} \Big( \left(\nicefrac{\Delta \rho}{\rho}\right)_{xy} + \left(\nicefrac{\Delta \rho}{\rho}\right)_{yx} \Big) \sin(2\theta) \\
\left(\nicefrac{\Delta \rho}{\rho}\right)_{(yy)''}(H_z) &= \left(\nicefrac{\Delta \rho}{\rho}\right)_{xx}\sin^2\theta + \left(\nicefrac{\Delta \rho}{\rho}\right)_{yy}\cos^2\theta - \frac{1}{2} \Big( \left(\nicefrac{\Delta \rho}{\rho}\right)_{xy} + \left(\nicefrac{\Delta \rho}{\rho}\right)_{yx} \Big) \sin(2\theta) \\
\left(\nicefrac{\Delta \rho}{\rho}\right)_{(xy)''}(H_z) &= \left(\nicefrac{\Delta \rho}{\rho}\right)_{xy}\cos^2\theta - \left(\nicefrac{\Delta \rho}{\rho}\right)_{yx}\sin^2\theta - \frac{1}{2} \Big( \left(\nicefrac{\Delta \rho}{\rho}\right)_{xx} - \left(\nicefrac{\Delta \rho}{\rho}\right)_{yy} \Big) \sin(2\theta) \\
\left(\nicefrac{\Delta \rho}{\rho}\right)_{(yx)''}(H_z) &= -\left(\nicefrac{\Delta \rho}{\rho}\right)_{xy}\sin^2\theta + \left(\nicefrac{\Delta \rho}{\rho}\right)_{yx}\cos^2\theta - \frac{1}{2} \Big( \left(\nicefrac{\Delta \rho}{\rho}\right)_{xx} - \left(\nicefrac{\Delta \rho}{\rho}\right)_{yy} \Big) \sin(2\theta).
\end{align}
\end{subequations}

\noindent The form of \eqref{eq:a4} emphasizes that the amount of $\left(\nicefrac{\Delta \rho}{\rho}\right)_{xy}$ and $\left(\nicefrac{\Delta \rho}{\rho}\right)_{yx}$ signal in a $\left(\nicefrac{\Delta \rho}{\rho}\right)_{(xx)''}$ or $\left(\nicefrac{\Delta \rho}{\rho}\right)_{(yy)''}$ measurement is characterized by $\sin(2\theta)$, and likewise for the amount of $\left(\nicefrac{\Delta \rho}{\rho}\right)_{(xx)}$ and $\left(\nicefrac{\Delta \rho}{\rho}\right)_{(yy)}$ in a $\left(\nicefrac{\Delta \rho}{\rho}\right)_{(xy)''}$ or $\left(\nicefrac{\Delta \rho}{\rho}\right)_{(yx)''}$ measurement.

For the in-plane transport measurements that are the subject of this work, the transformation properties of the relevant elastoresistivity coefficients are given by performing the transformation \eqref{eq:5} with the specific elastoresistivity tensor \eqref{eq:a1}:

\begin{subequations}
\label{eq:a5}
\begin{align}
m_{(xx)'',(xx)'} &= m_{(yy)'',(yy)'} =  m_{xx,xx} - m_{xy,xy} \sin(2\theta) \sin(2\phi) - \tfrac{1}{2} \left( m_{xx,xx} - m_{xx,yy} \right) \Big[ 1 - \cos(2\theta) \cos(2\phi) \Big] \\
m_{(xx)'',(yy)'} &= m_{(yy)'',(xx)'} = m_{xx,yy} + m_{xy,xy} \sin(2\theta) \sin(2\phi) + \tfrac{1}{2} \left( m_{xx,xx} - m_{xx,yy} \right) \Big[ 1 - \cos(2\theta) \cos(2\phi) \Big] \\
m_{(xx)'',(zz)'} &= m_{(yy)'',(zz)'} = m_{xx,zz} \\
m_{(xy)'',(xx)'} &= m_{xy,xx} - m_{xy,xy} \cos(2\theta) \sin(2\phi) - \tfrac{1}{2} (m_{xx,xx} - m_{xx,yy}) \sin(2\theta) \cos(2\phi) \\
m_{(xy)'',(yy)'} &= m_{xy,xx} + m_{xy,xy} \cos(2\theta) \sin(2\phi) + \tfrac{1}{2} (m_{xx,xx} - m_{xx,yy}) \sin(2\theta) \cos(2\phi) \\
m_{(yx)'',(xx)'} &= -m_{xy,xx} - m_{xy,xy} \cos(2\theta) \sin(2\phi) - \tfrac{1}{2} (m_{xx,xx} - m_{xx,yy}) \sin(2\theta) \cos(2\phi) \\
m_{(yx)'',(yy)'} &= -m_{xy,xx} + m_{xy,xy} \cos(2\theta) \sin(2\phi) + \tfrac{1}{2} (m_{xx,xx} - m_{xx,yy}) \sin(2\theta) \cos(2\phi) \\
m_{(xy)'',(zz)'} &= -m_{(yx)'',(zz)'} = m_{xy,zz}.
\end{align}
\end{subequations}

Expressing the in-plane transport quantities $\left(\nicefrac{\Delta \rho}{\rho}\right)_{(ij)''}$ in terms of applied normal strains $\epsilon_{(kl)'}$, plugging in the transformed elastoresistivity coefficients from \eqref{eq:a5}, and taking the symmetry-motivated combinations $\left(\nicefrac{\Delta \rho}{\rho}\right)_{(xx)''} - \left(\nicefrac{\Delta \rho}{\rho}\right)_{(yy)''}$ and $\left(\nicefrac{\Delta \rho}{\rho}\right)_{(xy)''} + \left(\nicefrac{\Delta \rho}{\rho}\right)_{(yx)''}$, we arrive at expressions \eqref{eq:6} and \eqref{eq:9} of the main text.

As a final comment, and in order to connect with the experimental setup described in our previous work,\cite{chu_2012,kuo_2013} we note that for configurations in which the current is sourced coincidentally with the normal strain axes, $\phi = -\theta$ and

\begin{equation}
\label{eq:a6}
m_{(ij)'',(kl)'} = \hat{\alpha}_{\theta} m_{ij,kl} \hat{\alpha}_{-\theta} = \hat{\alpha}_{\theta} m_{ij,kl} \hat{\alpha}^{-1}_{\theta}.
\end{equation}

\section{Quantifying Current and Strain Alignment Errors}\label{appendix:alignmenterrors}

As described in the main text (and illustrated in Figures \ref{fig:symmetryconfiga}, \ref{fig:symmetryconfigb}), the high-symmetry configurations for the differential longitudinal elastoresistance measurement are given by \eqref{eq:6} with $(\theta, \phi) = (0,0) \textrm{ or } (-\pi/4,\pi/4)$, while the high-symmetry configurations for the superposed transverse elastoresistance measurment are given by \eqref{eq:9} with $(\theta, \phi) = (\pi/4,0) \textrm{ or } (0,\pi/4)$.  In an actual measurement, however, slight misalignment relative to these high symmetry directions can be anticipated.  In this section we quantify the consequences of such misalignments.

Suppose that in a measurement of $\left(\nicefrac{\Delta \rho}{\rho}\right)_{(xx)''}$, it was intended that the crystal frame be oriented relative to the transport and normal strain frames according to some $(\theta, \phi)$, but the actual configuration was given by $(\theta + \delta \theta_{xx}, \phi + \delta \phi_{xx})$.  Similarly, suppose that in an attempt to measure $\left(\nicefrac{\Delta \rho}{\rho}\right)_{(yy)''}$, the intended orientation of the crystal frame to be oriented relative to the transport and normal strain frames was $(\theta, \phi)$, but the actual configuration was given by $(\theta + \delta \theta_{yy}, \phi + \delta \phi_{yy})$.  In general, we take $\delta \theta_{xx} \neq \delta \theta_{yy}$ and $\delta \phi_{xx} \neq \delta \phi_{yy}$ so that we treat all alignment errors independently. We now show how such errors propagate in the experimental determination of the relevant elastoresistivity coefficients.

For this type of misalignment, the combination $\left(\nicefrac{\Delta \rho}{\rho}\right)_{(xx)''}(H_z) - \left(\nicefrac{\Delta \rho}{\rho}\right)_{(yy)''}(H_z)$ is given by

\begin{align}
\label{eq:b1}
\left(\nicefrac{\Delta \rho}{\rho}\right)&_{(xx)''} - \left(\nicefrac{\Delta \rho}{\rho}\right)_{(yy)''} = -\frac{1}{2} (\epsilon_{(xx)'} - \epsilon_{(yy)'}) 2m_{xy,xy} \Big[ \sin2(\theta + \delta \theta_{xx})\sin2(\phi + \delta \phi_{xx}) + \sin2(\theta + \delta \theta_{yy})\sin2(\phi + \delta \phi_{yy}) \Big] \nonumber \\
& + \frac{1}{2} (\epsilon_{(xx)'} - \epsilon_{(yy)'}) (m_{xx,xx} - m_{xx,yy}) \Big[ \cos2(\theta + \delta \theta_{xx})\cos2(\phi + \delta \phi_{xx}) + \cos2(\theta + \delta \theta_{yy})\cos2(\phi + \delta \phi_{yy}) \Big],
\end{align}

\noindent This expression naturally reduces to \eqref{eq:6} for perfect angular alignment (i.e., $\delta \theta_{xx} = \delta \theta_{yy} = \delta \phi_{xx} = \delta \theta_{yy} = 0$).  Expanding this expression to quadratic order in the angular errors about the high-symmetry configurations $(\theta, \phi) = (0,0) \textrm{ and } (\theta, \phi) = (-\pi/4,\pi/4)$, we find

\begin{subequations}
\label{eq:b2}
\begin{align}
\left(\nicefrac{\Delta \rho}{\rho}\right)&_{(xx)''} - \left(\nicefrac{\Delta \rho}{\rho}\right)_{(yy)''} \approx -(\epsilon_{(xx)'} - \epsilon_{(yy)'}) 2m_{xy,xy} \Big[ 2 (\delta \theta_{xx} \delta \phi_{xx} + \delta \theta_{yy} \delta \phi_{yy}) \Big] \\
&+ (\epsilon_{(xx)'} - \epsilon_{(yy)'}) (m_{xx,xx} - m_{xx,yy}) \Big[ 1 - ( (\delta \theta_{xx})^2 + (\delta \theta_{yy})^2 + (\delta \phi_{xx})^2 + (\delta \phi_{yy})^2 ) \Big] \qquad \qquad \bm{\Big[ (\theta, \phi) = (0,0) \Big]} \nonumber
\end{align}

\begin{align}
\left(\nicefrac{\Delta \rho}{\rho}\right)&_{(xx)''} - \left(\nicefrac{\Delta \rho}{\rho}\right)_{(yy)''} \approx (\epsilon_{(xx)'} - \epsilon_{(yy)'}) 2m_{xy,xy} \Big[ 1 - ( (\delta \theta_{xx})^2 + (\delta \theta_{yy})^2 + (\delta \phi_{xx})^2 + (\delta \phi_{yy})^2 ) \Big] \\
&- (\epsilon_{(xx)'} - \epsilon_{(yy)'}) (m_{xx,xx} - m_{xx,yy}) \Big[ 2 (\delta \theta_{xx} \delta \phi_{xx} + \delta \theta_{yy} \delta \phi_{yy}) \Big]. \qquad \qquad \qquad \qquad \bm{\Big[ (\theta, \phi) = (-\pi/4,\pi/4) \Big]} \nonumber
\end{align}

\end{subequations}

\noindent As can be seen from \eqref{eq:b2}, for the high-symmetry configurations, the amount of error introduced enters at second order in the angular misalignments.

Consideration of alignment errors for the transverse elastoresistance configuration proceeds in an analogous manner.  Suppose that for a high-symmetry configuration $(\theta,\phi)$, the measurements of $\left(\nicefrac{\Delta \rho}{\rho}\right)_{(xy)''}$ and $\left(\nicefrac{\Delta \rho}{\rho}\right)_{(yx)''}$ are actually characterized by $(\theta + \delta \theta_{xy},\phi + \delta \phi_{xy})$ and $(\theta + \delta \theta_{yx},\phi + \delta \phi_{yx})$, respectively, with $\delta \theta_{xy} \neq \delta \theta_{yx}$ and $\delta \phi_{xy} \neq \delta \phi_{yx}$.  A measurement of $\left(\nicefrac{\Delta \rho}{\rho}\right)_{(xy)''}(H_z) + \left(\nicefrac{\Delta \rho}{\rho}\right)_{(yx)''}(H_z)$ is then given by

\begin{align}
\label{eq:b3}
\left(\nicefrac{\Delta \rho}{\rho}\right)&_{(xy)''} + \left(\nicefrac{\Delta \rho}{\rho}\right)_{(yx)''} = -\frac{1}{2} (\epsilon_{(xx)'} - \epsilon_{(yy)'}) 2m_{xy,xy} \Big[ \cos2(\theta + \delta \theta_{xy})\sin2(\phi + \delta \phi_{xy}) + \cos2(\theta + \delta \theta_{yx})\sin2(\phi + \delta \phi_{yx}) \Big] \nonumber \\
& -\frac{1}{2} (\epsilon_{(xx)'} - \epsilon_{(yy)'}) (m_{xx,xx} - m_{xx,yy}) \Big[ \sin2(\theta + \delta \theta_{xy})\cos2(\phi + \delta \phi_{xy}) + \sin2(\theta + \delta \theta_{yx})\cos2(\phi + \delta \phi_{yx}) \Big].
\end{align}

\noindent which reduces to \eqref{eq:9} for perfect angular alignment.  Again expanding this expression to quadratic order in the angular errors about the high-symmetry configurations $(\theta, \phi) = (\pi/4,0) \textrm{ and } (\theta, \phi) = (0,\pi/4)$, we find

\begin{subequations}
\label{eq:b4}
\begin{align}
\left(\nicefrac{\Delta \rho}{\rho}\right)&_{(xy)''} + \left(\nicefrac{\Delta \rho}{\rho}\right)_{(yx)''} \approx (\epsilon_{(xx)'} - \epsilon_{(yy)'}) 2m_{xy,xy} \Big[ 2 (\delta \theta_{xy} \delta \phi_{xy} + \delta \theta_{yx} \delta \phi_{yx}) \Big] \\
&- (\epsilon_{(xx)'} - \epsilon_{(yy)'}) (m_{xx,xx} - m_{xx,yy}) \Big[ 1 - ( (\delta \theta_{xy})^2 + (\delta \theta_{yx})^2 + (\delta \phi_{xy})^2 + (\delta \phi_{yx})^2 ) \Big] \qquad \qquad \bm{\Big[ (\theta, \phi) = (\pi/4,0) \Big]} \nonumber
\end{align}

\begin{align}
\left(\nicefrac{\Delta \rho}{\rho}\right)&_{(xy)''} + \left(\nicefrac{\Delta \rho}{\rho}\right)_{(yx)''} \approx -(\epsilon_{(xx)'} - \epsilon_{(yy)'}) 2m_{xy,xy} \Big[ 1 - ( (\delta \theta_{xy})^2 + (\delta \theta_{yx})^2 + (\delta \phi_{xy})^2 + (\delta \phi_{yx})^2 ) \Big] \\
&+ (\epsilon_{(xx)'} - \epsilon_{(yy)'}) (m_{xx,xx} - m_{xx,yy}) \Big[ 2 (\delta \theta_{xy} \delta \phi_{xy} + \delta \theta_{yx} \delta \phi_{yx}) \Big]. \qquad \qquad \qquad \qquad \qquad \qquad \bm{\Big[ (\theta, \phi) = (0,\pi/4) \Big]} \nonumber
\end{align}
\end{subequations}

\noindent Once again, the errors enter at second order about the high-symmetry configurations.

\section{Quantifying Strain Magnitude Errors}\label{appendix:strainmagnitudeerrors}

A second type of error to consider is the case of a differential or superposed elastoresistance measurement in which the two samples are perfectly aligned but experience unequal strains.  We do not make any specific assumptions about the physical origin of this difference, which might be different depending on the specific experimental configuration that is chosen.  For the specific technique that we have used in which thin crystals are adhered to the side surface of a piezoelectric stack, such a difference can arise from imperfect strain transmission by the epoxy used to adhere the crystals to the stack or from strain relaxation due to geometric considerations.

Suppose that an experiment is characterized by a fixed $(\theta,\phi)$ configuration but that different relative amounts of strain are experienced by the two samples during a differential longitudinal elastoresistance measurement (i.e., the strains which induce a finite $(\nicefrac{\Delta \rho}{\rho})_{(xx)''}$ are slightly different than those causing $(\nicefrac{\Delta \rho}{\rho})_{(yy)''}$). In other words, $(\nicefrac{\Delta \rho}{\rho})_{(xx)''}$ is measured in response to a strain $\epsilon_{(kl)'}^{(1)}$ and $(\nicefrac{\Delta \rho}{\rho})_{(yy)''}$ is measured in response to a slightly different strain $\epsilon_{(kl)'}^{(2)}$, with $\epsilon_{(kl)'}^{(1)} \textrm{ and } \epsilon_{(kl)'}^{(2)}$ given by

\begin{equation}
\label{eq:c1}
\epsilon_{(kl)'}^{(1)} = \begin{pmatrix} \epsilon_{(xx)'} + \delta \epsilon_{(xx)'}^{(1)} \\ \epsilon_{(yy)'} + \delta \epsilon_{(yy)'}^{(1)} \\ \epsilon_{(zz)'} + \delta \epsilon_{(zz)'}^{(1)} \\ 0 \\ 0 \\ 0 \\ 0 \\ 0 \\ 0 \end{pmatrix} \quad \textrm{ and } \quad \epsilon_{(kl)'}^{(2)} = \begin{pmatrix} \epsilon_{(xx)'} + \delta \epsilon_{(xx)'}^{(2)} \\ \epsilon_{(yy)'} + \delta \epsilon_{(yy)'}^{(2)} \\ \epsilon_{(zz)'} + \delta \epsilon_{(zz)'}^{(2)} \\ 0 \\ 0 \\ 0 \\ 0 \\ 0 \\ 0 \end{pmatrix}.
\end{equation}

\noindent The associated resistivity changes are given by

\begin{subequations}
\label{eq:c2}
\begin{align}
(\nicefrac{\Delta \rho}{\rho})_{(xx)''} &= m_{(xx)'',(xx)'}\epsilon_{(xx)'}^{(1)} + m_{(xx)'',(yy)'}\epsilon_{(yy)'}^{(1)} + m_{(xx)'',(zz)'}\epsilon_{(zz)'}^{(1)} \\
(\nicefrac{\Delta \rho}{\rho})_{(yy)''} &= m_{(yy)'',(xx)'}\epsilon_{(xx)'}^{(2)} + m_{(yy)'',(yy)'}\epsilon_{(yy)'}^{(2)} + m_{(yy)'',(zz)'}\epsilon_{(zz)'}^{(2)}
\end{align}
\end{subequations}

\noindent with the individual elastoresistivity coefficients transforming according to \eqref{eq:a5}.  Explicitly incorporating the angular dependence of the $m_{(ij)'',(kl)'}$, the symmetric combination $\left(\nicefrac{\Delta \rho}{\rho}\right)_{(xx)''}(H_z)-\left(\nicefrac{\Delta \rho}{\rho}\right)_{(yy)''}(H_z)$ is given as

\begin{align}
\label{eq:c3}
\left(\nicefrac{\Delta \rho}{\rho}\right)_{(xx)''} -& \left(\nicefrac{\Delta \rho}{\rho}\right)_{(yy)''} = (\epsilon_{(xx)'}-\epsilon_{(yy)'}) \Big[ (m_{xx,xx} - m_{xx,yy})\cos2\theta \cos2\phi - 2m_{xy,xy}\sin2\theta \sin2\phi \Big] \\
&+ \frac{1}{2}(\delta \epsilon_{(xx)'}^{(1)} + \delta \epsilon_{(xx)'}^{(2)} - \delta \epsilon_{(yy)'}^{(1)} - \delta \epsilon_{(yy)'}^{(2)}) \Big[ (m_{xx,xx} - m_{xx,yy})\cos2\theta \cos2\phi - 2m_{xy,xy}\sin2\theta \sin2\phi \Big] \nonumber \\
&+ \frac{1}{2} (\delta \epsilon_{(xx)'}^{(1)} - \delta \epsilon_{(xx)'}^{(2)} + \delta \epsilon_{(yy)'}^{(1)} - \delta \epsilon_{(yy)'}^{(2)}) \Big[ m_{xx,xx} + m_{xx,yy} \Big] + (\delta \epsilon_{(zz)'}^{(1)} - \delta \epsilon_{(zz)'}^{(2)})\Big[ m_{xx,zz} \Big] \nonumber,
\end{align}

\noindent where we have organized each term based on the particular irrep of $D_{4h}$ to which it corresponds.  The leading term in \eqref{eq:c3} is simply \eqref{eq:6} (i.e., the combination of $B_{1g}$ and $B_{2g}$ quantities dictated by the specific $(\theta,\phi)$ configuration), but we also measure error terms in proportion to the amount of $\delta \epsilon_{_{A_{1g,1}}} \equiv \frac{1}{2} (\delta \epsilon_{(xx)'}^{(1)} - \delta \epsilon_{(xx)'}^{(2)} + \delta \epsilon_{(yy)'}^{(1)} - \delta \epsilon_{(yy)'}^{(2)})$, $\delta \epsilon_{_{A_{1g,2}}} \equiv \delta \epsilon_{(zz)'}^{(1)} - \delta \epsilon_{(zz)'}^{(2)}$, and $\delta \epsilon_{_{B_{1g}}} \equiv \frac{1}{2}(\delta \epsilon_{(xx)'}^{(1)} + \delta \epsilon_{(xx)'}^{(2)} - \delta \epsilon_{(yy)'}^{(1)} - \delta \epsilon_{(yy)'}^{(2)})$ strain that is also applied during the measurement.  Note that the degree of error in the differential elastoresistance measurement is at first order in the magnitude of the relative strain offset.

An analogous expression can be worked out for the superposed transverse elastoresistance configuration as well.  Suppose that $(\nicefrac{\Delta \rho}{\rho})_{(xy)''}$ is measured in response to a strain $\epsilon_{(kl)'}^{(3)}$ and $(\nicefrac{\Delta \rho}{\rho})_{(yx)''}$ is measured in response to a slightly different strain $\epsilon_{(kl)'}^{(4)}$, with $\epsilon_{(kl)'}^{(3)} \textrm{ and } \epsilon_{(kl)'}^{(4)}$ given by

\begin{equation}
\label{eq:c4}
\epsilon_{(kl)'}^{(3)} = \begin{pmatrix} \epsilon_{(xx)'} + \delta \epsilon_{(xx)'}^{(3)} \\ \epsilon_{(yy)'} + \delta \epsilon_{(yy)'}^{(3)} \\ \epsilon_{(zz)'} + \delta \epsilon_{(zz)'}^{(3)} \\ 0 \\ 0 \\ 0 \\ 0 \\ 0 \\ 0 \end{pmatrix} \quad \textrm{ and } \quad \epsilon_{(kl)'}^{(4)} = \begin{pmatrix} \epsilon_{(xx)'} + \delta \epsilon_{(xx)'}^{(4)} \\ \epsilon_{(yy)'} + \delta \epsilon_{(yy)'}^{(4)} \\ \epsilon_{(zz)'} + \delta \epsilon_{(zz)'}^{(4)} \\ 0 \\ 0 \\ 0 \\ 0 \\ 0 \\ 0 \end{pmatrix}.
\end{equation}

\noindent The corresponding changes in resistivity are then

\begin{subequations}
\label{eq:c5}
\begin{align}
(\nicefrac{\Delta \rho}{\rho})_{(xy)''} &= m_{(xy)'',(xx)'}\epsilon_{(xx)'}^{(3)} + m_{(xy)'',(yy)'}\epsilon_{(yy)'}^{(3)} + m_{(xy)'',(zz)'}\epsilon_{(zz)'}^{(3)} \\
(\nicefrac{\Delta \rho}{\rho})_{(yx)''} &= m_{(yx)'',(xx)'}\epsilon_{(xx)'}^{(4)} + m_{(yx)'',(yy)'}\epsilon_{(yy)'}^{(4)} + m_{(yx)'',(zz)'}\epsilon_{(zz)'}^{(4)}
\end{align}
\end{subequations}

\noindent with the individual elastoresistivity coefficients transforming according to \eqref{eq:a5}.  Explicitly incorporating the angular dependence of the $m_{(ij)'',(kl)'}$, the symmetric combination $\left(\nicefrac{\Delta \rho}{\rho}\right)_{(xy)''}(H_z)+\left(\nicefrac{\Delta \rho}{\rho}\right)_{(yx)''}(H_z)$ is given as

\begin{align}
\label{eq:c6}
\left(\nicefrac{\Delta \rho}{\rho}\right)_{(xy)''} +& \left(\nicefrac{\Delta \rho}{\rho}\right)_{(yx)''} = -(\epsilon_{(xx)'} - \epsilon_{(yy)'}) \Big[ (m_{xx,xx} - m_{xx,yy})\cos2\theta \sin2\phi + 2m_{xy,xy}\sin2\theta \cos2\phi \Big] \\
&- \frac{1}{2}(\delta \epsilon_{(xx)'}^{(3)} + \delta \epsilon_{(xx)'}^{(4)} - \delta \epsilon_{(yy)'}^{(3)} - \delta \epsilon_{(yy)'}^{(4)}) \Big[ (m_{xx,xx} - m_{xx,yy})\cos2\theta \sin2\phi + 2m_{xy,xy}\sin2\theta \cos2\phi \Big] \nonumber \\
&+ \frac{1}{2}(\delta \epsilon_{(xx)'}^{(3)} - \delta \epsilon_{(xx)'}^{(4)} + \delta \epsilon_{(yy)'}^{(3)} - \delta \epsilon_{(yy)'}^{(4)}) \Big[ m_{xy,xx} \Big] + (\delta \epsilon_{(zz)'}^{(3)} - \delta \epsilon_{(zz)'}^{(4)})\Big[ m_{xy,zz} \Big] \nonumber.
\end{align}

\noindent Just as in the differential longitudinal case, the measured errors in the transverse superposed elastoresistance configuration appear in proportion to the amount of $\delta \epsilon_{_{A_{1g,1}}} \equiv \frac{1}{2}(\delta \epsilon_{(xx)'}^{(3)} - \delta \epsilon_{(xx)'}^{(4)} + \delta \epsilon_{(yy)'}^{(3)} - \delta \epsilon_{(yy)'}^{(4)})$, $\delta \epsilon_{_{A_{1g,2}}} \equiv \delta \epsilon_{(zz)'}^{(3)} - \delta \epsilon_{(zz)'}^{(4)}$, and $\delta \epsilon_{_{B_{1g}}} \equiv \frac{1}{2}(\delta \epsilon_{(xx)'}^{(3)} + \delta \epsilon_{(xx)'}^{(4)} - \delta \epsilon_{(yy)'}^{(3)} - \delta \epsilon_{(yy)'}^{(4)})$ strain that are inadvertently applied during the measurement, and the degree of error is at first order in the magnitude of the relative strain offset.  However, since $\left(\nicefrac{\Delta \rho}{\rho}\right)_{(yx)''}(H_z) = \left(\nicefrac{\Delta \rho}{\rho}\right)_{(xy)''}(-H_z)$, one can constrain these strain offsets to be precisely zero since the measurements can be performed under the same mounting conditions (only inverting the magnetic field environment).  That is, for a measurement performed on one single crystal, $\delta \epsilon_{(xx)'}^{(4)} = \delta \epsilon_{(xx)'}^{(3)}$ and $\delta \epsilon_{(yy)'}^{(4)} = \delta \epsilon_{(yy)'}^{(3)}$, such that $\delta \epsilon_{_{A_{1g,2}}} = 0$.  Rephrased in terms of group theory, the strain error does not mix symmetry channels (a measurement of the $B_{1g}$ response is not contaminated by any $A_{1g}$ signal), but the absolute magnitude of the strain experienced by the sample is incorrectly recorded as $\epsilon_{(xx)'}$ rather than $\epsilon_{(xx)'} + \delta \epsilon_{(xx)'}^{(3)}$ and $\epsilon_{(yy)'}$ rather than $\epsilon_{(yy)'} + \delta \epsilon_{(yy)'}^{(4)}$.  Thus, provided one can isolate $(\nicefrac{\Delta \rho}{\rho})_{xy}$ and $(\nicefrac{\Delta \rho}{\rho})_{yx}$ (i.e., subtract out $\rho_{xx}$ contributions which ``contaminate'' a nominally $\rho_{xy}$ measurement, which we describe in the following Appendix \ref{appendix:transverseisolation}), the transverse configuration is immune to contamination from other symmetry channels arising from relative strain magnitude errors.  As discussed in the main text, this is the primary advantage of the transverse elastoresistance technique.

\section{Isolating Transverse Elastoresistivities in a Longitudinally-Contaminated Elastoresistance Measurement}\label{appendix:transverseisolation}

In this section we describe in detail the process by which a longitudinal resistivity ($\rho_{xx}$) ``contamination'' can be subtracted from an erstwhile transverse resistivity ($\rho_{xy}$) measurement. Such a situation is often encountered in the context of Hall effect measurements, for which a nominally transverse signal $R_{xy}^{\textrm{measured}}$ (due to a $\hat{z}$-oriented magnetic field $\boldsymbol H = H_z \hat{z}$) is contaminated by a contribution from the longitudinal resistance $R_{xx}$ such that

\begin{equation}
\label{eq:d1}
R_{xy}^{\textrm{measured}} = R_{xy} + \alpha R_{xx},
\end{equation}

\noindent where $\alpha$ is a (not generally small) parameter characterizing the amount of $R_{xx}$ contamination. Such a contamination can arise, for example, from curved current paths within the crystal or from imperfectly aligned voltage contacts.  In the Hall effect measurement case, to isolate the Hall resistance signal, one typically measures $R_{xy}^{\textrm{measured}}$ for both positive and negative magnetic field and uses the fact that the longitudinal (transverse) signal is even (odd) in $H_z$:

\begin{equation}
\label{eq:d2}
R_{xy}^{\textrm{measured}}(H_z) - R_{xy}^{\textrm{measured}}(-H_z) = 2R_{xy}(H_z).
\end{equation}

\noindent In this appendix, we will describe the analog to this magnetic field anti-symmetrization for a transverse elastoresistance measurement.  In contrast to the Hall effect, though, we cannot rely on a simple parity-in-field argument since contamination of a nominal $2m_{xy,xy}$ or $m_{xx,xx}-m_{xx,yy}$ measurement can come from elastoresistivity coefficients that are either even (e.g., $m_{xx,zz}$) or odd (e.g., $m_{xy,xx}$) in the magnetic field.  Nevertheless, by pre-characterizing the contact geometry, it is still possible to isolate the symmetry-connected elastoresistivity coefficients which are related to thermodynamic susceptibilities.

\begin{figure}
\includegraphics[trim = 0mm 0mm 0mm 0mm, clip, scale=1]{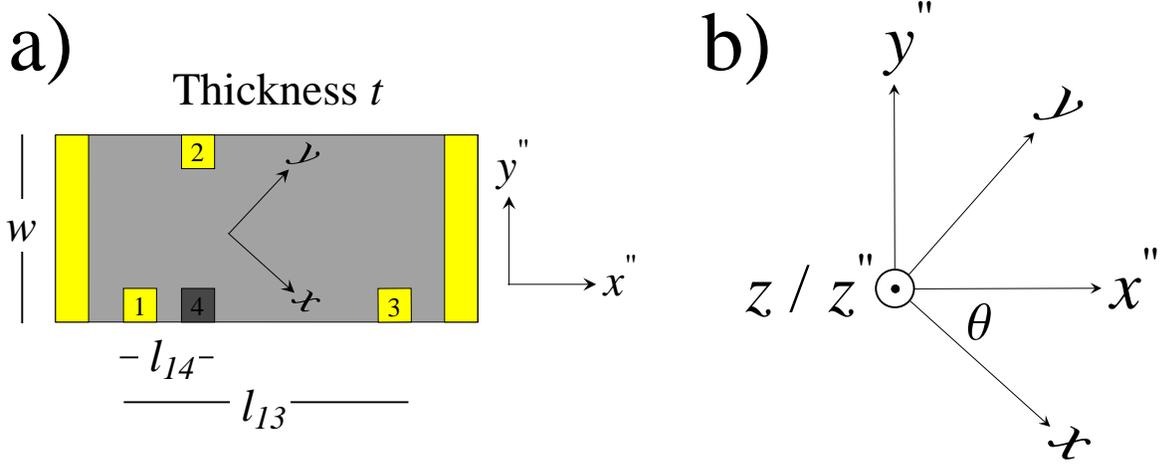}
\centering
\caption{(a) Schematic diagram illustrating contact misalignment in a transverse measurement.  The transverse voltage is to be measured between contacts 1 and 2, while the longitudinal voltage can be measured between contacts 1 and 3. The degree of transverse contact misalignment (horizontal offset between contacts 1 and 2) has been greatly exaggerated for pedagogical purposes.  Contact 4 (dark gray) is a hypothetical contact which is perfectly vertically aligned with contact 2, which means that the relative voltages between these contacts $V_2 - V_4 = 0$ in zero magnetic field for zero strain. (b) Primitive crystal frame and its relative alignment to the transport frame.}
\label{fig:transversemisalignment}
\end{figure}

In performing a transverse resistivity measurement, the most straightforward manner in which one might encounter longitudinal contamination is due to imperfect contact geometry (as in Figure~\ref{fig:transversemisalignment}) or to nonlinearly directed current paths within the crystal.  While microscopic details to do with the specific mechanism of current non-uniformity would dictate how such a longitudinal contamination would be subtracted out, we instead focus on how to subtract out the longitudinal contamination due to contact misalignment, otherwise assuming a homogeneous material with uniformly directed current paths.

Consider an experimental configuration as in Figure~\ref{fig:transversemisalignment} in which one sources a current $\vec{I} = I \cdot \hat{x}'' = [ w t j_{(x)''} ] \cdot \hat{x}''$ along the $x''$ direction in the transport frame (with $j_{(x)''}$ the magnitude of the current density in the transport frame, $w$ the crystal width, and $t$ the crystal thickness). One then seeks to isolate the true transverse resistivity $\rho_{(xy)''}$ (in the transport frame) from a measurement $R_{(xy)''}^{\textrm{measured}}$ which contains longitudinal contamination $\rho_{(xx)''}$ due to imperfectly aligned contacts 1 and 2.  One most conveniently characterizes this contamination by writing Ohm's law in the transport frame as

\begin{gather}
\label{eq:d3}
E_{(i)''} = \rho_{(ij)''} j_{(j)''} \\
\begin{pmatrix} E_{(x)''} \\[3pt] E_{(y)''} \\[3pt] E_{(z)''} \end{pmatrix} 
= \begin{pmatrix} \rho_{(xx)''} & \rho_{(xy)''} & 0 \\[3pt] \rho_{(yx)''} & \rho_{(yy)''} & 0 \\[3pt] 0 & 0 & \rho_{(zz)''} \end{pmatrix} \begin{pmatrix} j_{(x)''} \\[3pt] 0 \\[3pt] 0 \end{pmatrix} = j_{(x)''} 
\begin{pmatrix} \rho_{(xx)''} \\[3pt] \rho_{(yx)''} \\[3pt] 0 \end{pmatrix} \nonumber
\end{gather}

\noindent and then expressing the resistances in terms of measured voltages, uniform electric fields, and crystal dimensions:

\begin{gather}
\label{eq:d4}
R^{\textrm{measured}}_{(xx)''} = \frac{V_3- V_1}{I_{(x)''}} = \frac{E_{(x)''}l_{13}}{I_{(x)''}} = \frac{l_{13}}{l_{14}} \frac{V_4 - V_1}{I_{(x)''}} \equiv \frac{1}{\Delta_{\ell}} \frac{V_4 - V_1}{I_{(x)''}} \\
R^{\textrm{measured}}_{(yx)''} = \frac{V_2 - V_1}{I_{(x)''}} = \frac{V_2 - V_4}{I_{(x)''}} + \frac{V_4 - V_1}{I_{(x)''}} = \frac{E_{(y)''} w}{j_{(x)''} w t} + \Delta_{\ell} R^{\textrm{measured}}_{(xx)''}  = \frac{\rho_{(yx)''} }{t} + \Delta_{\ell} R^{\textrm{measured}}_{(xx)''} \nonumber,
\end{gather}

\noindent where $V_i$ denotes a voltage measured at the $i$-th contact in Figure \ref{fig:transversemisalignment} and $\Delta_{\ell} \equiv \frac{l_{14}}{l_{13}}$ characterizes the degree of misalignment of the transverse contacts 1 and 2.  Solving for $\rho_{(yx)''}$ in \eqref{eq:d4}, one obtains

\begin{equation}
\label{eq:d5}
\rho_{(yx)''} = t\left( R^{\textrm{measured}}_{(yx)''} - \Delta_{\ell} R^{\textrm{measured}}_{(xx)''} \right) \equiv t R_{(yx)''},
\end{equation}

\noindent where $R_{(yx)''}$ represents the resistance that would be measured in the absence of contact misalignment.  Isolating $R_{(yx)''}$ thus requires down-weighting the simultaneously measured quantity $R^{\textrm{measured}}_{(xx)''}$ by the parameter $\Delta_{\ell}$, which is most readily determined via a zero-field, zero-strain measurement of the contacts.  Specifically, in the absence of either strain or a magnetic field, there cannot be a transverse voltage for a crystal with $D_{4h}$ symmetry; therefore, $V_2 = V_4$ and, with the assumption of a uniform electric field within the material,

\begin{equation}
\label{eq:d6}
\Delta_{\ell} \equiv \frac{l_{14}}{l_{13}} = \frac{V_4 - V_1}{V_3 - V_1} \: \stackrel{\hat{\epsilon}=\hat{0}, H_z = 0}{=} \: \frac{V_2 - V_1}{V_3 - V_1}.
\end{equation}

The equations \eqref{eq:d5} and \eqref{eq:d6} express how one corrects for contact misalignment in order to isolate the resistance $R_{(yx)''}$; however, as prescribed in \eqref{eq:4} of the main text, the meaningful quantity for elastoresistivity is the ratio of \textit{resistivities}, not resistances.  The relationship between the two is elucidated by considering the transverse quantities $\left( \nicefrac{\Delta \rho}{\rho} \right)_{(yx)''} (H_z)$ and $\left( \nicefrac{\Delta \rho}{\rho} \right)_{(xy)''} (H_z)$, where

\begin{gather}
\label{eq:d7}
\left( \frac{\Delta \rho}{\rho} \right)_{(yx)''} = \frac{\Delta\rho_{(yx)''}}{\sqrt{\rho_{(xx)''}(\hat{\epsilon}=\hat{0}) \rho_{(yy)''}(\hat{\epsilon}=\hat{0})}} \\
\left( \frac{\Delta \rho}{\rho} \right)_{(xy)''} = \frac{\Delta\rho_{(xy)''}}{\sqrt{\rho_{(xx)''}(\hat{\epsilon}=\hat{0}) \rho_{(yy)''}(\hat{\epsilon}=\hat{0})}} \nonumber
\end{gather}

\noindent and where the normalization factor contains the unstrained \textit{longitudinal} resistivities $\rho_{(xx)''}$ and $\rho_{(yy)''}$. Since $\rho_{(yx)''} = tR_{(yx)''}$ (and likewise for $\rho_{(xy)''}$), the linearized strain-induced resistivity change is given by $\Delta \rho_{(yx)''} = t \Delta R_{(yx)''} + R_{(yx)''} \Delta t$; substituting into \eqref{eq:d7}, and using the fact that for a tetragonal material $\rho_{(xx)''}(\hat{\epsilon}=\hat{0}) = \rho_{(yy)''}(\hat{\epsilon}=\hat{0})$,

\begin{equation}
\label{eq:d8}
\left( \frac{\Delta \rho}{\rho} \right)_{(yx)''} = \frac{t \Delta R_{(yx)''} + R_{(yx)''} \Delta t}{\left( \frac{wt}{l_{13}} \right) R_{(xx)''}(\hat{\epsilon}=\hat{0})},
\end{equation}

\noindent and likewise for $\left( \frac{\Delta \rho}{\rho} \right)_{(xy)''}$.  When the strain is of a $B_{1g}$ or $B_{2g}$ character (i.e., area-preserving), the thickness change due to strain $\Delta t$ is precisely zero and so the second term in the numerator of \eqref{eq:d8} vanishes (i.e., $R_{(yx)''} \Delta t = 0$).  When the strain is predominantly of a $B_{1g}$ or $B_{2g}$ character but also possesses an area-deforming ($A_{1g}$) component (as is a more realistic approximation to our specific experimental realization), $\Delta t$ is finite; nevertheless, provided the measured resistance change is dominated by changes in the resistivity under strain as opposed to geometric effects, it will still be true that $R_{(yx)''} \Delta t \ll t \Delta R_{(yx)''}$.  In either case then, $\Delta \rho_{(yx)''} \approx t \Delta R_{(yx)''}$, and, combining \eqref{eq:d5} and \eqref{eq:d8},

\begin{equation}
\label{eq:d9}
\left( \frac{\Delta \rho}{\rho} \right)_{(yx)''} \approx \frac{l_{13}}{w} \bigg[ \left( \frac{\Delta R_{(yx)''}^{\textrm{measured}}}{R_{(xx)''}} \right) - \Delta_{\ell} \left( \frac{\Delta R_{(xx)''}^{\textrm{measured}}}{R_{(xx)''}} \right) \bigg],
\end{equation}

\noindent and likewise for $\left( \frac{\Delta \rho}{\rho} \right)_{(xy)''}$.  In \eqref{eq:d9}, it is to be understood that the approximation becomes an exact equality under conditions of pure $B_{1g}$ or $B_{2g}$ strain.  This is the procedure we used to subtract longitudinal resistance ``contamination'' and obtain the data described in the main text.

In zero magnetic field and for appropriate mounting configurations, a single measurement with contact misalignment accounted for according to \eqref{eq:d9} is sufficient for extracting the relevant elastoresistivity coefficients; in a finite field, one requires an extra measurement that is performed after reversing the magnetic field.  Upon taking the symmetry-motivated sum $\left( \nicefrac{\Delta \rho}{\rho} \right)_{(yx)''} (H_z) + \left( \nicefrac{\Delta \rho}{\rho} \right)_{(xy)''} (H_z)$ (and using \eqref{eq:d9} and the Onsager relations), the generalized expression for finite $H_z$ is given by

\begin{gather}
\label{eq:d10}
\left( \frac{\Delta \rho}{\rho} \right)_{(yx)''} \hspace{-7mm} (H_z)+\left( \frac{\Delta \rho}{\rho} \right)_{(xy)''} \hspace{-7mm} (H_z) = \left( \frac{\Delta \rho}{\rho} \right)_{(yx)''} \hspace{-7mm} (H_z) + \left( \frac{\Delta \rho}{\rho} \right)_{(yx)''} \hspace{-7mm} (-H_z) \\
= \frac{l_{13}}{w} \bigg[ \left( \frac{\Delta R_{(yx)''}^{\textrm{measured}}(H_z)}{R_{(xx)''}(H_z)} \right) - \Delta_{\ell} \left( \frac{\Delta R_{(xx)''}^{\textrm{measured}}(H_z)}{R_{(xx)''}(H_z)} \right) + \left( \frac{\Delta R_{(yx)''}^{\textrm{measured}}(-H_z)}{R_{(xx)''}(-H_z)} \right) - \Delta_{\ell} \left( \frac{\Delta R_{(xx)''}^{\textrm{measured}}(-H_z)}{R_{(xx)''}(-H_z)} \right) \bigg]. \nonumber
\end{gather}

\noindent Equations \eqref{eq:d9} and \eqref{eq:d10} express the transverse elastoresistivity analog to anti-symmetrization in a magnetic field for Hall resistance measurements.  By measuring $\left( \nicefrac{\Delta R_{(yx)''}^{\textrm{measured}}}{R_{(xx)''}} \right)$ and $\left( \nicefrac{\Delta R_{(xx)''}^{\textrm{measured}}}{R_{(xx)''}} \right)$ simultaneously, and having pre-characterized $\Delta_{\ell}$ in zero magnetic field (using \eqref{eq:d6}, measured under conditions of zero strain), one can simply subtract the two elastoresistance measurements (with the longitudinal contribution weighted by $\Delta_{\ell}$ and an overall geometric correction by $\frac{l_{13}}{w}$) in order to isolate $\left( \nicefrac{\Delta \rho}{\rho} \right)_{(yx)''}$.  The same subtraction procedure works for measuring either $m_{xx,xx} - m_{xx,yy}$ or $2m_{xy,xy}$ since the above derivation is independent of the relative orientation of the transport and crystal frames.

\section{Evaluating Goodness of Fit of Curie-Weiss Model to the Measured Elastoresistivity Coefficient $2m_{xy,xy}$}\label{appendix:cwparameterfit}

The elastoresistivity coefficient $2m_{xy,xy}$ exhibits a monotonic increase with decreasing temperature from the highest temperature measured (220 K in the present work) down to 136 K, which is just above the coupled structural and magnetic transition temperature $T_{S,N} \approx 134$ K. In this section, we briefly describe the procedures used to fit the data to a Curie-Weiss temperature dependence above the phase transition, which is physically motivated based on a mean-field description of the nematic susceptibility.\cite{chu_2012,kuo_2013}

\begin{figure}
\includegraphics[trim = 0mm 0mm 0mm 0mm, clip, scale=1]{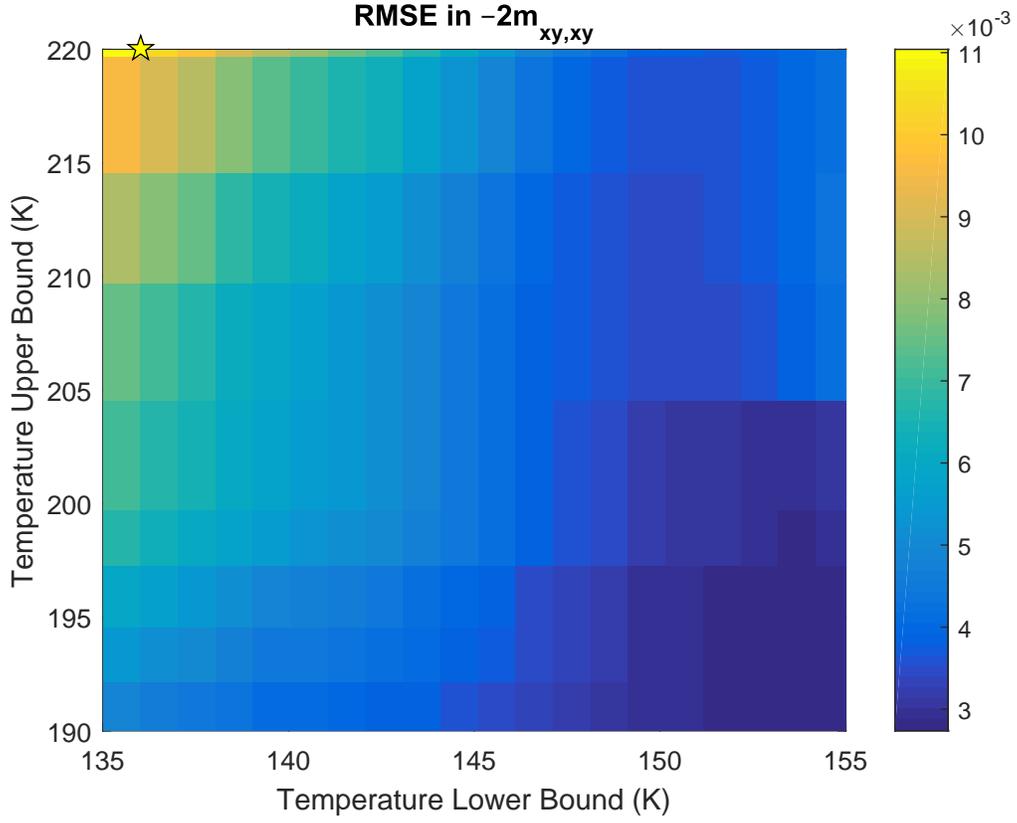}
\centering
\caption{Bullseye plot of the $2m_{xy,xy}$ elastoresistivity coefficient, illustrating the minimized RMSE from the Curie-Weiss form as a function of a varying temperature window. While an ideal fit over a particular range would produce a local minimum in the RMSE in that range and hence a bullseye-like pattern in the contour map, no such bullseye pattern is observed for the present data. This indicates that there is no ``optimal'' temperature range over which the data best conforms to the Curie-Weiss form.  Without such a best-fit range, we opt to fit over the maximal range from 220 K (the highest temperature measured) down to 136 K (just above $T_{S,N} \approx 134$ K).}
\label{fig:RMSE}
\end{figure}

As discussed in the main text, the $2m_{xy,xy}$ elastoresistivity data were fit to a Curie-Weiss temperature dependence of the form $2m_{xy,xy}^{\textrm{Curie-Weiss}} = \frac{\lambda}{a_0(T-\theta)} + 2m_{xy,xy}^0$ (see \eqref{eq:12} of the main text).  In order to evaluate the goodness of fit of the Curie-Weiss form to the measured data, we illustrate in Figure~\ref{fig:RMSE} a bullseye plot as a function of a varying temperature window.  A bullseye plot is a contour map of the root mean square error (RMSE) of the observed data from an expected model as a function of a varying window in the independent variable.  In the ideal case where the expected model perfectly conforms to the measured data over a particular range, the RMSE will obtain a local minimum over this range and the output of the contour map will resemble a bullseye-like pattern. Since narrowing the independent variable window also diminishes the sample size over which the fit is performed and relaxes constraints on the parameters in the fit model, bullseye plots also display a general trend of decreasing RMSE as one increases (decreases) the low (high) end cutoff in the independent variable window; one therefore expects a general trend of decreasing RMSE as one approaches the bottom right region of the contour map.

In the context of the present measurement, the expected model is a Curie-Weiss temperature dependence and the color output of the bullseye plot in Figure~\ref{fig:RMSE} corresponds to the residual difference (for each temperature window) between the inverse susceptibility $(2m_{xy,xy}-2m_{xy,xy}^0)^{-1}$ and a linear function of temperature.  For each fixed window, we first estimate the temperature-independent parameter $2m_{xy,xy}^0$ by a least RMSE minimization procedure between the measured $2m_{xy,xy}$ and the Curie-Weiss form; with this $2m_{xy,xy}^0$, we then estimate the best-fit parameters $\nicefrac{\lambda}{a_0}$ and $\theta$ by minimizing the RMSE between the $(2m_{xy,xy}-2m_{xy,xy}^0)^{-1}$ and a linear fit.  The low temperature cutoff for the windows varies between 135 K and 155 K, while the high temperature cutoff for the windows varies between 190 K and 220 K.  As displayed in Figure~\ref{fig:RMSE}, a bullseye-like pattern is not observed; the only discernible feature is a general trend of decreasing RMSE as the temperature window is narrowed. This indicates the absence of an ``optimal'' temperature range over which the elastoresistivity coefficient $2m_{xy,xy}$ displays Curie-Weiss behavior; therefore, we choose to fit the data over the maximal range from 220 K (the highest temperature measured) down to 136 K (just above $T_{S,N}$).  This range is demarcated by a star in the top-left portion of Figure~\ref{fig:RMSE} and yields the fit parameters given in the main text (see Table~\ref{table1}).

\begin{figure}
\includegraphics[trim = 0mm 0mm 0mm 0mm, clip, scale=1]{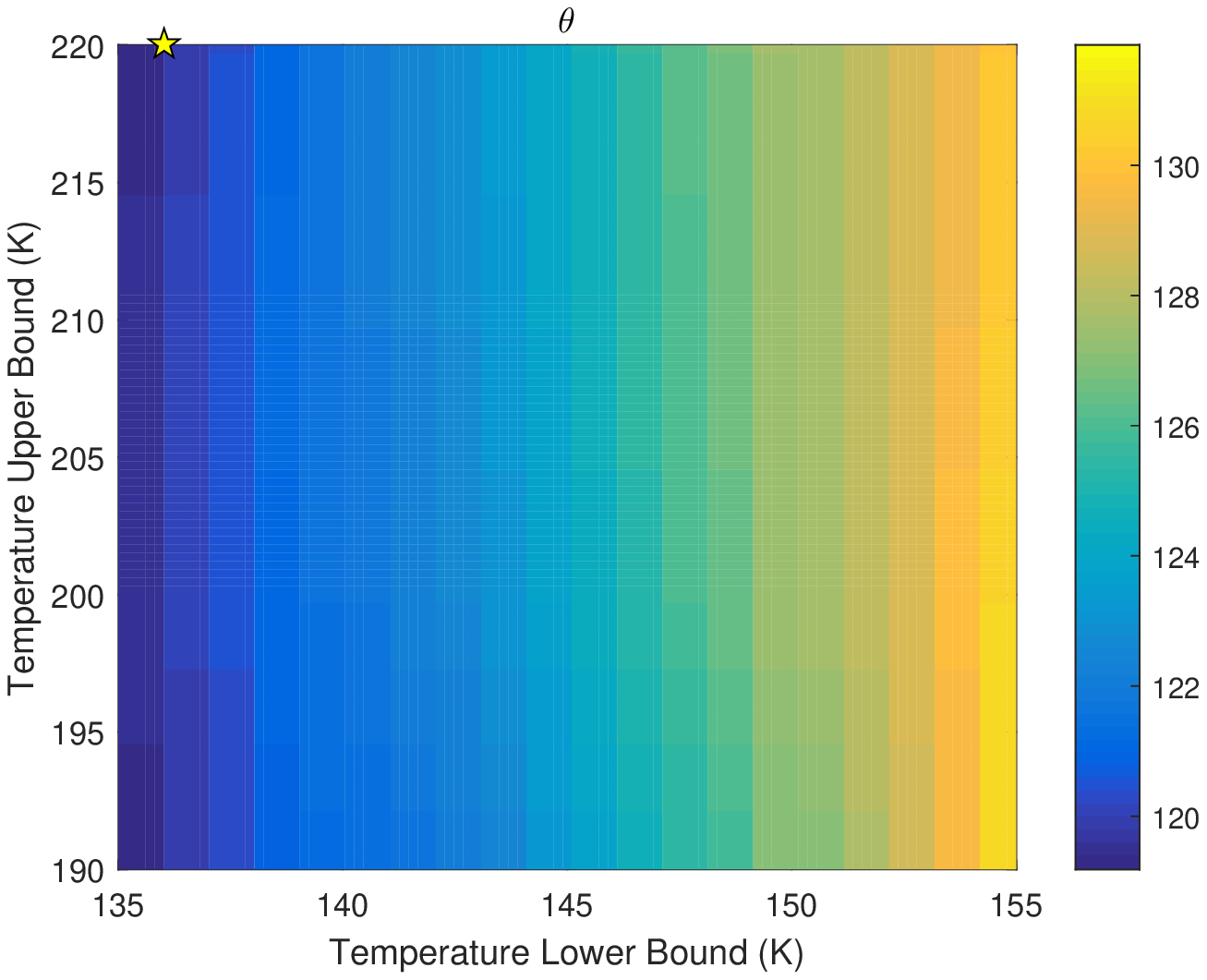}
\centering
\caption{Contour map of estimated Weiss temperatures $\theta$ as a function of a varying temperature window.  The star in the top-left region again corresponds to the maximal temperature range, which we use due to the absence of any particular best fit range.  The distribution of Weiss temperatures is heavily concentrated around $\sim 120$ K for the ranges close to the maximal window, which is the value of $\theta$ quoted in the main text.}
\label{fig:weisstemperaturefits}
\end{figure}

The most easily interpretable parameter from the Curie-Weiss fit is the Weiss temperature $\theta$.  To characterize the dependence of the Weiss temperature on the particular temperature window used for the fit, we also plot a contour map of the best-fit estimates of $\theta$ as a function of a varying temperature window (Figure~\ref{fig:weisstemperaturefits}).  The star in the top-left region of Figure~\ref{fig:weisstemperaturefits} again corresponds to the maximal temperature range, which we use due to the absence of any particular best fit range.  The distribution of Weiss temperatures is heavily concentrated around $\sim 120$ K for the ranges close to the maximal window, which is the value of $\theta$ quoted in the main text.

\begin{thebibliography}{99}


\bibitem{kuo_2013} Hsueh-Hui Kuo, Maxwell C. Shapiro, Scott C. Riggs, and Ian R. Fisher, Phys. Rev. B \textbf{88}, 085113 (2013).

\bibitem{shapiro_2015} M. C. Shapiro, Patrik Hlobil, A. T. Hristov, Akash V. Maharaj, and I. R. Fisher, Phys. Rev. B \textbf{92}, 235147 (2015).

\bibitem{chu_2012} Jiun-Haw Chu, Hsueh-Hui Kuo, James G. Analytis, and Ian R. Fisher, Science \textbf{337}, 710 (2012).

\bibitem{kuo_2014} Hsueh-Hui Kuo and Ian R. Fisher, Phys. Rev. Lett. \textbf{112}, 227001 (2014).

\bibitem{riggs_2015} Scott C. Riggs, M. C. Shapiro,	Akash V. Maharaj,	S. Raghu,	E. D. Bauer, R. E. Baumbach, P. Giraldo-Gallo, Mark Wartenbe, and I. R. Fisher, Nature Comm. \textbf{6}, 6425 (2015).

\bibitem{hlobil_2015} Patrik Hlobil, Akash V. Maharaj, Pavan Hosur, M. C. Shapiro, I. R. Fisher, and S. Raghu, Phys. Rev. B \textbf{92}, 035148 (2015).

\bibitem{sun_2010} Y. Sun, S. Thompson, and T. Nishida, \textit{Strain Effect in Semiconductors: Theory and Device Applications} (Springer, New York, 2010).

\bibitem{watson_2015} M. D. Watson, T. K. Kim, A. A. Haghighirad, N. R. Davies, A. McCollam, A. Narayanan, S. F. Blake, Y. L. Chen, S. Ghannadzadeh, A. J. Schofield, M. Hoesch, C. Meingast, T. Wolf, and A. I. Coldea, Phys. Rev. B \textbf{91}, 155106 (2015).

\bibitem{kuo_2015} Hsueh-Hui Kuo, Jiun-Haw Chu, Steven A. Kivelson, and Ian R. Fisher, arXiv:1503.00402v1 (2015).

\bibitem{tanatar_2015} M. A. Tanatar, A. E. B\"{o}hmer, E. I. Timmons, M. Sch\"{u}tt, G. Drachuck, V. Taufour, S. L. Bud'ko, P. C. Canfield, R. M. Fernandes, R. Prozorov, arXiv:1511.04757 (2015).

\bibitem{kuo_2016} Hsueh-Hui Kuo, Jiun-Haw Chu, J. C. Palmstrom, Steven A. Kivelson, and Ian R. Fisher, Unpublished (2016).

\bibitem{onsager_1931} L. Onsager, Phys. Rev. \textbf{38}, 2265 (1931).

\bibitem{footnote1} For our initial elastoresistance measurements on the iron-based superconductors, $A_{1g}$ strain contamination does not affect any of the conclusions since the observation of a singular divergence in $\chi_{_{B_{2g}}}$ is immune to imperfect $A_{1g}$ subtraction.  A divergence in the $A_{1g}$ symmetry channel would manifest equally in both $\chi_{_{B_{1g}}}$ and $\chi_{_{B_{2g}}}$ (which is not observed), and so an imperfect $A_{1g}$ subtraction at worst contributes a small, non-singular, temperature-dependent background.

\bibitem{footnote2} On cooling from room temperature to 250 K, the PZT expands by $\sim 0.02 \%$ along its poling direction,\cite{simpson_1987} which is opposite to the typical thermal contraction of most materials.  Silver (a represenative material) and BaFe$_2$As$_2$ (the test material used in this manuscript) contract by $\sim 0.08 \%$ and $\sim 0.05 \%$ on cooling to 250 K, respectively,\cite{touloukian_1975,bud'ko_2009} and so by adhering either material to a PZT, the materials are placed under uniaxial tension of $\sim 0.1 \%$ and $\sim 0.07 \%$, respectively.  These strains are smaller than the strain change of $\sim 0.15 \%$ that is obtained by sweeping the applied voltage to the PZT between $-$50 V to +150 V at 250 K.\cite{kuo_2013,piezo_data_sheet}

\bibitem{simpson_1987} A. M. Simpson and W. Wolfs, Rev. Sci. Instrum. \textbf{58}, 2193 (1987).

\bibitem{touloukian_1975} Touloukian, Y. S., Kirby, R. K., Taylor R. E., and Desai, P. D., eds., Thermophysical Properties of Matter: The TPRC Data Series, Vol. 12, \textit{Thermal Expansion: Metallic Elements and Alloys}, Plenum, NY (1975) p. 298$-$299.

\bibitem{bud'ko_2009} S. L. Bud'ko, N. Ni, S. Nandi, G. M. Schmiedeshoff, and P. C. Canfield, Phys. Rev. B \textbf{79}, 054525 (2009).

\bibitem{hicks_2014} Clifford W. Hicks, Mark E. Barber, Stephen D. Edkins, Daniel O. Brodsky, and Andrew P. Mackenzie, Rev. Sci. Instrum. \textbf{85}, 065003 (2014).

\bibitem{piezo_data_sheet} Physik Instrumente GmbH, ``Piezo Material Data''.

\bibitem{sefat_2008} A. S. Sefat, R. Jin, M. A. McGuire, B. C. Sales, D. J. Singh, and D. Mandrus, Appl. Phys. Lett. \textbf{101}, 117004 (2008).

\bibitem{chu_2009} J.-H. Chu, J. G. Analytis, C. Kucharczyk, and I. R. Fisher, Phys. Rev. B \textbf{79}, 014506 (2009).

\bibitem{footnote3} Temperature stability is an important consideration for these measurements, with the degree of stability needed dictated by the particular material at hand.  A requisite condition is that the strain-induced response $\Delta \rho(T,\epsilon+\Delta\epsilon)$ be much larger than the thermally-induced resistivity change $\Delta \rho(T+\Delta T,\epsilon)$; equivalently, this condition requires $\Delta T \ll \nicefrac{\frac{d\rho}{d\epsilon} \cdot \Delta \epsilon}{\frac{d\rho}{dT}}$, where $\frac{d\rho}{d\epsilon}$ is related to an appropriate combination of the material's elastoresistivity coefficients.

\bibitem{man_2015} Haoran Man, Xingye Lu, Justin S. Chen, Rui Zhang, Wenliang Zhang, Huiqian Luo, J. Kulda, A. Ivanov, T. Keller, Emilia Morosan, Qimiao Si, and Pengcheng Dai, Phys. Rev. B \textbf{92}, 134521 (2015).

\bibitem{kuo_thesis} Hsueh-Hui Kuo, ``Electronic Nematicity in Iron-Based Superconductors'', Ph.D. Thesis, Stanford University (2014).

\bibitem{fernandes_2014} R. M. Fernandes,	A. V. Chubukov, and J. Schmalian, Nature Phys. \textbf{10}, 97 (2014).


\end{thebibliography}
\end{document}